\documentclass[11pt]{article}
\pdfoutput=1 
\usepackage{jheppub}
\usepackage{tabu}
\usepackage[vcentermath]{youngtab}
\usepackage[usenames,dvipsnames,table]{xcolor}
\usepackage{graphicx,amsmath,amssymb,amsthm,multirow,array,bm,bbm,esint}
\usepackage[mathscr]{eucal}
\usepackage[bbgreekl]{mathbbol}
\usepackage{epsf,amsfonts}
\usepackage{slashed}
\usepackage[numbers,sort&compress]{natbib}
\usepackage{minibox}
\usepackage{rotating}
\usepackage{pdflscape}

\newcommand{\be}{\begin{equation}} 
\newcommand{\ee}{\end{equation}}

\usepackage{upgreek}

\definecolor{Orange}{rgb}{1,.4,0}

\hypersetup{
    pdfstartview={FitH},    
    pdftitle={EFT for Chaos in CFTs},    
    pdfauthor={Felix Haehl and Moshe Rozali},     
    colorlinks=true,       
    linkcolor=blue,          
    citecolor=blue,        
    filecolor=blue,      
    urlcolor=blue           
}

\title{Effective Field Theory for Chaotic CFTs}

\author{Felix M.\ Haehl,}
\author{Moshe Rozali}

\affiliation{
\it{
Department of Physics and Astronomy, University of British Columbia,\\
Vancouver, BC V6T 1Z1, Canada}}

\emailAdd{f.m.haehl@gmail.com}
\emailAdd{rozali@phas.ubc.ca}

\abstract{We derive an effective field theory for general chaotic two-dimensional conformal field theories with a large central charge. The theory is a specific and calculable instance of a more general framework recently proposed in [1]. We discuss the gauge symmetries of the model and how they relate to the Lyapunov behaviour of certain correlators. We calculate the out-of-time-ordered correlators diagnosing quantum chaos, as well as certain more fine-grained higher-point generalizations, using our Lorentzian effective field theory. We comment on potential future applications of the effective theory to real-time thermal physics and conformal field theory.}

\begin{document} 
 
\maketitle
\flushbottom

\section{Introduction}

Work on quantum gravity is increasingly informed by methods and considerations of quantum many-body physics, as holography relates quantum gravity and black holes to a more conventional boundary description. Gravity seems to emerge as a collective description of certain chaotic many-body systems. Indeed it seems to be distinguished by being maximally chaotic, at least with respect to some measures of (early time) chaos \cite{Maldacena:2015waa}. 

Perhaps more puzzling is an apparent connection, revealed in several  models, between early time chaos, as quantified by the quantum Lyapunov exponent \cite{Larkin:1969aa,Shenker:2013pqa,Shenker:2013yza,Leichenauer:2014nxa,Maldacena:2015waa,Kitaev:2015aa}, and the late time physics of diffusion and transport (e.g. \cite{Gu:2016oyy,Patel:2016wdy,Swingle:2016jdj,Blake:2016wvh,Blake:2016jnn,Blake:2016sud,Blake:2017qgd,Davison:2016ngz}). Such a connection has motivated the authors of \cite{Blake:2017ris} to propose an effective action to account for the effect of the early time chaos. 
In that effective action,  the initial exponential growth of out-of-time-order correlators (OTOCs) arises from the exchange of an effective mode. Furthermore, at least in the maximally chaotic (i.e. near-coherent) case the proposed effective mode corresponds to energy fluctuations. In other words the effective theory is very similar to hydrodynamics, albeit used in an unusual regime where gradients are not necessarily small.

This effective action provides a generalization of the Schwarzian action describing $AdS_2$ gravity and the low frequency physics of the SYK model \cite{Kitaev:2015aa,Maldacena:2016hyu,Kitaev:2017awl,Mertens:2017mtv,Engelsoy:2016xyb,Turiaci:2017zwd}. In that context, a connection to hydrodynamics was proposed early on \cite{Jensen:2016pah}. Crucially, the SYK model has an emergent conformal symmetry, which is broken by the Schwarzian effective action. In turn, the Schwarzian action has an $SL(2,\mathbb{R})$ gauge redundancy. The model of \cite{Blake:2017ris} more generally postulates an emergent shift symmetry which would be responsible both for the Lyapunov behaviour of OTOCs and for the absence of exponential growth in more conventional, time-ordered correlators. The origin of such a shift symmetry, as well as the identification of the effective scrambling mode,  remain to be derived from a more microscopic description.

It is useful then to have another controlled example (in addition to the SYK model) where the microscopic physics is relatively well-understood, and assumptions and extrapolations in the general case can be demonstrated. Further, it is desirable to find classes of models in higher dimensions exhibiting features similar to the SYK model. To address these issues, we derive an effective theory for chaotic two-dimensional conformal field theories (CFTs) with large central charge. 
While we shall indeed focus on measures of quantum chaos, our model is really an effective description of stress tensor exchanges. Under the assumption of vacuum dominance (which is underlying our discussion) the physics of stress tensor exchanges provides a universal sector of any 2D CFT and thus allows for a unified effective description.
We see that the general structure postulated in \cite{Blake:2017ris} holds in this specific case, and the microscopic origin of the scrambling mode and the shift symmetry are transparent. Both are intimately tied to the presence of conformal symmetry. 

The structure of our discussion is as follows. We start by demonstrating that a basic prediction of the effective models of \cite{Blake:2017ris}, the so-called ``pole skipping'' phenomenon (also seen in  the holographic context in \cite{Grozdanov:2017ajz}), holds in the context of chaotic two-dimensional CFTs with large central charge. This also  provides an extremely simple calculation of the Lyapunov exponent and of the butterfly velocity (previously calculated in \cite{Roberts:2014ifa}). 

We proceed by discussing the theory of the ``soft modes'' responsible for scrambling, which, in our context, are holomorphic and anti-holomorphic reparametrization modes, in section \ref{soft}. Our starting point is similar to that of \cite{Turiaci:2016cvo}, where it was shown that the Lyapunov growth of OTOCs in rational large-$c$ CFTs can be derived by thinking about the conformal transformations as physical Goldstone modes. However, for the purpose of calculating correlators, we compute the Lorentzian propagators of the soft modes. In section \ref{prop} we discuss the detailed structure of the theory (to quadratic order) on the Schwinger-Keldysh contour, as well as more complicated ``higher-OTO'' contours (with more switchbacks in time) which are needed for our purpose. This discussion may be of independent interest for real time conformal field theory.

We discuss in section \ref{external} the coupling between external probes (taken to be primary fields, for simplicity) and the soft modes. Using the ingredients discussed thus far, and assuming dominance of the soft mode for stress tensor exchanges, we calculate in section \ref{correlation} the correlation functions expressing the chaotic behaviour of two-dimensional CFTs with large central charge. We discuss first the 4-point function responsible for the Lyapunov behaviour and contrast it with the time-ordered correlators. We then calculate the higher-point extensions defined in \cite{Haehl:2017pak}, which measure more fine-grained notions of scrambling, and comment on the results.

We conclude by outlining directions for future research, including possible applications of the effective theory of the soft modes of 2D CFTs to physics unrelated to chaos. Appendix \ref{app:euclidean} contains some detailed Euclidean calculations and extensions. In Appendix \ref{sl2} we review details on the $SL(2,\mathbb{R})$ gauge symmetry of the soft mode action and its microscopic origin.\\

{\bf Note added:} While this paper was being finalized, we learned about \cite{Cotler:2018xxx}. Since there is some overlap with our discussion, we coordinated publication.

\vspace{10pt}
\section{Pole Skipping}
\label{pole}

The authors of \cite{Blake:2017ris} proposed an effective description of chaotic systems which encompasses the hydrodynamics, i.e., the theory of the energy-momentum tensor (and possibly other conserved currents), and quantum chaos as manifested in the out-of-time-order correlators (OTOCs). The theory differs from conventional hydrodynamics in having gradients of order unity. Thus the perturbative expansion has to be distinct from the usual gradient expansion, and is usually identified with an expansion in large $N$. We comment further on the validity of that expansion below.

As the effective description should be valid for all large $N$ chaotic systems, at least with maximal quantum Lyapunov exponent, we are motivated to explore the effective theory for chaotic CFTs with large central charge $c$. Before we discuss that formulation in the next section, we check that a distinct signature of such an effective description is realized in our chosen context.

Indeed, the effective description of  \cite{Blake:2017ris} predicts a certain ``pole skipping'' when considering retarded correlators of the energy-momentum tensor. In this section we verify that two-dimensional CFTs exhibit this phenomenon. Our aim is to calculate retarded correlators of the energy-momentum tensor, which factorize into holomorphic and anti-holomorphic sectors in two dimensions. We start by performing the calculation in  the holomorphic sector. 

On the complex plane, parameterized by the complex coordinate $z$, one has
\begin{equation}
\langle T(z_1) T(z_2) \rangle = \frac{c}{2(z_1-z_2)^4} \,.
\end{equation}
Transforming to the cylinder, $z=\exp(-i u)$, i.e., a finite temperature state with $\beta=2\pi$,
this gives
\begin{equation}
\langle T(u_1) T(u_2) \rangle_c = \frac{c}{32}\frac{1}{\sin (\frac{u_1-u_2}{2})^4} \,,
\end{equation}
where we have restricted attention to the connected 2-point function (subtracting off the disconnected part generated by the Schwarzian of the transformation between the plane and the cylinder).
We denote $u_1-u_2 = u = \tau+i\sigma$ such that the above expression is translation invariant and periodic in Euclidean time $\tau \in [0,2\pi)$.

 The Euclidean correlator in momentum space is obtained via a Fourier transform
\begin{equation}
\label{pos}
G^E_{(T)}(\omega_E,k) \equiv \langle T(\omega_E, k) T(-\omega_E, -k) \rangle_c = \frac{c}{32}\int d\tau d\sigma\, \frac{e^{-i \omega_E \tau-i k \sigma}}{\sin (\frac{\tau+i \sigma}{2})^4} \,,
\end{equation}
where the Euclidean frequency takes integer values, $\omega_E \in \mathbb{Z}$, and will be analytically continued  at the end of the calculation to find the retarded two-point function. The Fourier transform gives (see Appendix \ref{app:fourier} for details):
\begin{equation}
\label{Euc}
\begin{split}
  G^E_{(T)}(\omega_E,k) 
  &= \frac{c\,\pi}{6}\, \frac{\omega_E(\omega_E^2-1)}{\omega_E+ik} \,.
 \end{split}
\end{equation}
 
While the philosophy of focusing on the soft mode is similar to hydrodynamics, we will see that the theory of the soft mode in our case is not dissipative, due to the special kinematics in two dimensions. Nevertheless, the effective theory of the soft mode gives rise to the same phenomenon pointed out in  \cite{Blake:2017ris}. Namely, moving along the lines of poles $\omega_E+i k=0 $, the pole is skipped at $\omega_E =1 =\frac{2 \pi}{\beta}$. This identifies the Lyapunov exponent as being maximal, and the butterfly velocity as being the speed of light. These are the results that were found previously (and more laboriously) by \cite{Roberts:2014ifa}. Note, however, that the pole skipping alone is not sufficient to conclude chaotic behaviour or determine the Lyapunov exponent: pole skipping is universal in all two-dimensional CFTs (including theories with non-maximal chaos, or irrational CFTs). Having a maximal Lyapunov exponent for the out-of-time-order correlator requires further assumptions (such as large central charge and vacuum block dominance as in the context of \cite{Roberts:2014ifa}).\footnote{ We thank Mukund Rangamani for conversations on this point.} We will later compute the Lyapunov exponent more carefully in the framework of our effective theory, clarifying the assumptions required.

\paragraph{Higher spin pole skipping:}$\!\!\!\!\!\!$\footnote{ We thank Kristan Jensen for conversations on this topic.}
We note in passing that the pole skipping as discussed above can also be observed for exchanges of higher spin currents. For example, if the theory contains a spin-3 current $W_3$, then we can Fourier transform the two-point function as before and find:
\begin{equation}
  \langle W_3(\omega_E,k) W_3(-\omega_E,-k) \rangle = \frac{5c}{384} \int d\tau d\sigma \, \frac{e^{-i\omega_E \tau - ik\sigma}}{\sin\left(\frac{\tau+i\sigma}{2}\right)^6}  = - \frac{c\pi}{72} \, \frac{\omega_E(\omega_E^2-1)(\omega_E^2-4)}{\omega_E+ik} \,.
\end{equation}
In addition to the pole skipping observed in the stress tensor correlator, this also skips poles at $\omega_E = \pm 2$, corresponding to the spin-3 Lyapunov exponent \cite{Perlmutter:2016pkf}. This should persist in a similar way for higher spins, and allows for incorporating such exchanges in our effective field theory. We will not pursue this further in the present paper, but it would be an interesting phenomenon to investigate.

\vspace{10pt}
\section{The Soft Modes: Euclidean Considerations}
\label{soft}

Motivated by the results in the previous section, we now formulate a theory of the soft modes for two-dimensional chaotic CFTs. While we are motivated by the discussion of \cite{Blake:2017ris}, our derivation is  different, making use of the special structure of two-dimensional conformal field theories. In the following sections we use this theory to discuss real time physics and signatures of chaos in that context.  

In the context of two-dimensional CFTs, both energy and momentum are conserved and the corresponding soft modes can be organized  as being holomorphic or anti-holomorphic reparametrization modes (see also  \cite{Turiaci:2016cvo} for a discussion of these Goldstone modes and their connection with quantum chaos). We consider the effective theory of those modes, at the quadratic level, at leading order in large central charge $c$. For simplicity we often discuss only the holomorphic sector, with the understanding that results for the anti-holomorphic sector are similar, and it is the combination of both that gives rise to physical results. 

The ``hydrodynamics'' of \cite{Blake:2017ris} (and \cite{Jensen:2016pah}) can be  considered as the theory of the soft modes, regarded as mappings from a fixed ``reference'' spacetime to a dynamical one. In modern language, fluid dynamics is a sigma model of the maps from the worldvolume spacetime into the physical spacetime \cite{Nickel:2010pr,Dubovsky:2011sj,Dubovsky:2011sk,Haehl:2015pja}. For path integrals with timefolded contours, more than one copy of the dynamical spacetime is needed.\footnote{  The need for a Schwinger-Keldysh description of hydrodynamic Goldstone modes is not tied to dissipation, but arises even for non-dissipative effects \cite{Haehl:2013hoa,Haehl:2015pja}.}  Additionally, we may consider the theory formulated in either the fixed or the dynamical spacetime. In the hydrodynamical context those descriptions are called the Lagrangian or Eulerian description of the fluid. We focus here on the description in terms of fixed spacetime. In the analogous context of $AdS_2$ gravity, this description is similar to the one  utilized in \cite{Maldacena:2016upp}, as opposed to \cite{Almheiri:2014cka}. Indeed, we will see that most calculations are in perfect analogy to those performed in \cite{Maldacena:2016upp} in the context of the SYK model and $AdS_2$ gravity.

Our soft mode action can be formulated in either Euclidean or Lorentzian signature. For the purposes of calculating the Lorentzian soft mode action, we discuss  two-dimensional conformal field theories at finite temperature, on the Schwinger-Keldysh contour. This entails doubling of the spacetime coordinates $(\tau^\alpha,\sigma^\alpha)$ where $\alpha= 1,2$ denote the two segments of the contour. Any conformal field such as the stress-energy tensor $T$ can be regarded as living on the contour, or alternatively there are two copies $T^\alpha(\tau,\sigma)$ living in the original spacetime.

\vspace{10pt}
\subsection{Quadratic Action for the Soft Modes}

Having dispensed with the preliminaries, we are ready to discuss the soft mode action. To obtain the effective action for the soft modes, we consider performing the CFT path integral in the presence of sources for conformal transformations $\delta z= \epsilon(z)$, which are generated by the currents $J(z)=\epsilon(z) T(z)$. Whereas such transformations are a global symmetry for holomorphic parameters $\epsilon(z)$, they become gapless modes for general transformation parameter $\epsilon(z,\bar{z})$.  That is, the action for the soft mode corresponding to holomorphic reparametrizations stems from its dependence on the anti-holomorphic coordinate $\bar{z}$. 

Those ``local'' conformal transformations are implemented, via the conformal Ward identities, by the addition of a term $\int d^2 z \,  \bar{\partial} (\epsilon^\alpha(z,\bar{z}) T^\alpha(z))$ to the Schwinger-Keldysh action. The results of that path integral define the effective action as $Z(\epsilon)=e^{i W(\epsilon)}$. We note that we are focussing here on infinitesimal conformal transformations: if we were to discuss instead finite conformal transformations, we would obtain a Schwarzian-like theory for the soft modes. As we are only interested in perturbation theory, the action for the infinitesimal perturbation of the soft modes is sufficient.
For our purposes we need to discuss the action to quadratic order (see, however, Appendix \ref{app:eucl3} for the third order generalization), and the coupling of the soft modes to external probes, which we do in turn in the next subsections.

An alternative picture of the soft-mode action is semi-holographic: we treat the external operators appearing in correlation functions as probes, and we are interested in the contribution to their correlation function due to interaction with the energy-momentum tensor of the large $c$ CFT. To sum up such contributions we can use standard ``external field'' methods, such as reviewed, for example, in \cite{Weinberg:1996kr}. While the soft mode $\epsilon$ starts its life as an external source, conjugate to the energy-momentum tensor, performing a Legendre transform trades energy-momentum fluctuations with fluctuations of the soft mode $\epsilon$. We end up calculating the contribution of soft mode fluctuations to correlators of the semi-holographic external probes.\footnote{  As the soft modes are elements of the metric, there is an obvious relation to previous discussions of 2D gravity (see, e.g., \cite{Polyakov:1987zb}). Another connection is the one with shadow representations \cite{SimmonsDuffin:2012uy}: analogous treatments for a conformal primary fields would make the ``external field'' transform in the shadow representation of the original primary. Treatment of the soft mode action beyond perturbation theory is beyond the scope of the present discussion.}

We note further that the mode $\epsilon(z,\bar z)$ formally has conformal weights $(h,\bar h) = (-1,0)$. This formal non-unitarity is the reason  $\epsilon$ will be able to exhibit an exponentially growing evolution. Such behaviour ultimately leads to the celebrated signature of chaos, i.e., exponentially growing out-of-time-order correlation functions.\\

We are interested in the effective action for the soft modes $\epsilon(z, \bar{z})$ and their anti-holomorphic counterparts, to quadratic order. Let us begin by working in Euclidean signature for simplicity.
From the above definition it is then clear that the quadratic term in the effective action is\footnote{  In some contexts (such as string theory) it might be conventional to rescale $\epsilon \rightarrow (2\pi)^{-1}\epsilon$.}
\begin{equation}
\label{quad}
W_2= - \frac{1}{2}\int d^2 z_1 \, d^2 z_2 \, \bar{\partial} \epsilon(z_1,\bar{z}_1) \,  \bar{\partial} \epsilon(z_2,\bar{z}_2) \, \langle T(z_1) T(z_2) \rangle_c + \text{(anti-holo.)}
\end{equation}
where we omit a similar expression for anti-holomorphic transformations, giving an additional (decoupled) soft modes $\bar{\epsilon}(z,\bar{z})$. 

So far, to be definite, we have used the notation where
our fixed coordinate system $(z, \bar{z})$ covers the complex plane, corresponding to the CFT at zero temperature. In that context our discussion is closely related to the induced gravity action of Polyakov \cite{Polyakov:1987zb}. However, an essential difference is that our soft mode describes the excitations of a thermal state (for example, this is the source of the pole skipping phenomena described above). Therefore,  henceforth we shall  work in cylinder coordinates $(u,\bar u)$ which describe a thermal state:
\begin{equation}
 z= e^{-iu} \equiv e^{-i(\tau+i\sigma)} \,, \qquad \bar{z} = e^{i\bar{u}}\equiv e^{i(\tau-i\sigma)}  \,.
\end{equation}
Explicitly, the quadratic action then reads as follows:
\begin{equation}
\label{eq:Wquad0}
\begin{split}
W_2 &= - \frac{c}{64}\int d^2u_1 d^2u_2 \,  \frac{\bar{\partial} \epsilon(u_1,\bar{u}_1) \, \bar{\partial}\epsilon(u_2 , \bar{u}_2) }{\sin^4 \left( \frac{u_1-u_2}{2} \right) } + \text{(anti-holo.)}  \,.
\end{split}
\end{equation}
Note that the quadratic action is of order $c$, thus in perturbative calculations each factor of the soft mode is accompanied by $1/\sqrt{c}$.

In Appendix \ref{app:eucl} we establish the Euclidean propagator of the soft mode from \eqref{eq:Wquad0}:
\begin{equation}\label{eq:epsProp}
\boxed{
\langle \epsilon(\omega_E,k) \, \epsilon (-\omega_E, -k) \rangle=  -\frac{24\pi}{c} \frac{1}{\omega_E\,(\omega_E^2-1) \, (\omega_E+ik) } \,.
}
\end{equation}
where $(\omega_E,k)$ are Euclidean momentum space coordinates related to the thermal coordinates $(\tau,\sigma)$ by a Fourier transform. 
The quadratic action for the soft mode (see \eqref{eq:WquadCalc}), which is quartic in derivatives, is very reminiscent of the (Euclidean) action of the SYK model, described in \cite{Maldacena:2016upp}, and the general (Lorentzian) action developed in \cite{{Blake:2017ris}}. Note that the spatial momentum appears only in the form of an anti-holomorphic derivative -- as expected a purely holomorphic $\epsilon(z) $ is a symmetry of the action, i.e. it  has zero action. A similar analysis for the anti-holomorphic mode yields
\begin{equation}\label{eq:epsPropBar}
\boxed{
\langle \bar\epsilon(\omega_E,k) \, \bar\epsilon (-\omega_E, -k) \rangle=  -\frac{24\pi}{c} \frac{1}{\omega_E\,(\omega_E^2-1) \, (\omega_E-ik) } \,,
}
\end{equation}
the only difference with respect to \eqref{eq:epsProp} being the sign of $k$.

By Fourier transforming the inverse of the above propagators (i.e., the quadratic action \eqref{eq:WquadCalc}), we can obtain the Euclidean quadratic action in thermal coordinates:
\begin{equation}
\boxed{
\begin{split}
W_2
%
&= \frac{c \pi}{6}  \int d\tau d\sigma \; \left[ \frac{1}{2}(\partial_\tau+i\partial_\sigma) \epsilon\;  (\partial_\tau^3 + \partial_\tau)  \epsilon  \;+\;  \frac{1}{2}  (\partial_\tau-i\partial_\sigma)\bar \epsilon\;  (\partial_\tau^3 + \partial_\tau)  \bar\epsilon \,  \right] \,,
\end{split}
}
\label{eq:WquadCalc2}
\end{equation}
where we added also the anti-holomorphic contribution. If the soft modes had been independent of $\sigma$, then the surviving terms in this action would be precisely two copies of the quadratic order piece in the expansion of the Schwarzian action in one dimension. The above procedure can be generalized to higher orders; see, for example, \eqref{eq:WcubicPos} for the analogous cubic action. 
Note that the action \eqref{eq:WquadCalc2} clearly has zero modes, a subset of which are associated with $SL(2,\mathbb{R})$ symmetries. We proceed to study these in more detail in the next subsection. 

We can add to the quadratic action a linear total derivative term for free. This corresponds to including a ground state energy. We write this suggestively as
\begin{equation}
\label{eq:W1}
\begin{split}
  W_1 
  &= \int d^2u \, \left[ \bar{\partial} \epsilon(u,\bar u) \, \langle T (u)\rangle + {\partial} \bar\epsilon(u,\bar u) \, \langle \bar{T} (\bar u) \rangle \right]=  \frac{c}{12} \int d\tau d\sigma  \left[ \frac{1}{2}(\partial_\tau + i \partial_\sigma) \epsilon + \frac{1}{2} (\partial_\tau - i \partial_\sigma) \bar \epsilon\right]
  \end{split}
\end{equation}
The normalization is such that it leads to the standard ground state energy density in the thermal state for two-dimensional CFTs: $\frac{E_0}{V} = \langle T(u) \rangle+\langle \bar{T}(\bar u)\rangle =  \frac{c}{12} \equiv  \frac{c}{12} (\frac{2\pi}{\beta})^2$.

\vspace{10pt}
\subsection{Noether Charges and $SL(2,\mathbb{R})$ Symmetries} 
\label{sec:noether}

We now investigate the quadratic action \eqref{eq:WquadCalc2}, and in particular its symmetries, in some more detail. The method and results will be similar to those used in the context of the Schwarzian quantum mechanics \cite{Maldacena:2016upp}, but the structure will be richer due to the extra dimension. 
Generically, the action \eqref{eq:WquadCalc2} has the following families of infinitesimal symmetries:\footnote{ We thank Kristan Jensen for comments on this point.}
\begin{equation}
   \delta_\text{h} \epsilon = \Lambda_\text{h}(\tau+i\sigma) \,,\qquad \delta_\pm \epsilon = \Lambda_\pm(\sigma)\, e^{\mp i \tau} \,,\qquad \delta_0 \epsilon = \Lambda_0(\sigma) \,,
\end{equation}
where $\Lambda_i$ are arbitrary functions. 
The associated on-shell conserved Noether currents are (up to the usual ambiguities)
\begin{equation}
\begin{split}
  J_\text{h}^\mu 
&= -\frac{\pi c}{6} \Big( \Lambda_\text{h} \, (\partial_\tau^3 + \partial_\tau)\epsilon \;,\; i \Lambda_\text{h} \, (\partial_\tau^3 + \partial_\tau)\epsilon \Big) \,,\\
  J_\pm^\mu &= -\frac{\pi c}{3} \Big( \Lambda_\pm\,e^{\mp i \tau} (\partial_\tau^2 \pm i \partial_\tau)  \bar\partial\epsilon \;, \;0  \Big) \,,\\
  J_0^\mu &= -\frac{\pi c}{3} \Big( \Lambda_0\,  (\partial_\tau^2 + 1)\bar\partial \epsilon \;, \; 0\Big)\,,
\end{split}
\end{equation}
where $J^\mu \equiv (J^\tau,J^\sigma)$. There are analogous expressions for the currents associated with the other reparametrization mode, $\bar\epsilon$. The first type of symmetry -- arbitrary infinitesimal holomorphic maps $\delta_\text{h}$ -- is one chiral half of the conformal invariance. The conservation of the current $J_\text{h}^\mu$, corresponding to general holomorphic transformations, can be interpreted as the conservation of the ``stress tensor'' current that couples to $\bar\partial \epsilon$ in \eqref{eq:WquadCalc2}:
\begin{equation}
\label{eq:noetherTh}
  \partial_\mu J^\mu_\text{h} = 2\,\bar{\partial} (\Lambda_\text{h}T_\text{h}) \simeq 0 \quad \text{ with } \quad T_\text{h} \equiv  -\frac{\pi c}{6}  \, (\partial_\tau^3 + \partial_\tau) \epsilon 
\end{equation}
where ``$\simeq$'' denotes use of the equations of motion.

Note that a naive dimensional reduction, which simply drops all $\sigma$-dependence, gives Noether charges that reproduce the linearized expressions in the context of $AdS_2$ gravity \cite{Maldacena:2016upp}. There, they were associated with the $SL(2,\mathbb{R})$ symmetry of the Schwarzian action.
This observation motivates us to highlight the transformations $\delta_{\pm,0}$ which are at the same time holomorphic functions of $\tau+i\sigma$. This amounts to setting 
\begin{equation}
  \Lambda_\pm(\sigma) = e^{\pm \sigma} \,,\qquad \Lambda_0(\sigma) = \text{const.}
\end{equation}
These transformations are symmetries of the Lagrangian in \eqref{eq:WquadCalc2} and correspond to the $SL(2,\mathbb{R})$ part of the Virasoro symmetry, which we review in more detail in Appendix \ref{sl2}. More precisely, the transformations correspond to chiral diffeomorphisms along the $SL(2,\mathbb{R})$ generating vector fields ${\cal L}_{\pm 1}$ and ${\cal L}_0$, defined in \eqref{eq:Lnvec}. 

We see that the pattern of symmetry breaking is analogous to that in the SYK model: the underlying Virasoro symmetry is explicitly broken by the conformal anomaly, and is spontaneously broken by the choice of a thermal background. It leaves behind the global conformal transformations $SL(2,\mathbb{R})$ which, upon Legendre transform to the soft mode description, are to be treated as gauge redundancies. This is crucial in ensuring that the exponentially growing mode is invisible in conventional correlators and thus does not represent a genuine instability of the system.

It is instructive to compute the zeroth order contribution to the charges, using the topological action \eqref{eq:W1}. Treating the constant shifts as a symmetry of that action, they give a contribution to $J_0^\tau$ (and to $\bar{J}_0^\tau$) which is $- \frac{c}{24}$. The associated Noether charges, obtained by integrating over the spatial direction, give a ground state entropy density, which reproduces the Cardy formula \cite{Cardy:1986ie}:
\begin{equation}
  -2\pi \big[Q_0 + \bar{Q}_0\big]_{{\cal O}(\epsilon^0)}  = -2\pi \times 2 \times  \int d\sigma\,\left(- \frac{c}{24} \right)  = \frac{c  \pi^2}{3} \frac{V}{\beta} = S_\text{Cardy} \,.
\end{equation}
where we reinstated $\beta$ and set formally $V = \int d\sigma$. The idea that entropy can be thought of as a Noether charge is very familiar in the context of black hole physics \cite{Iyer:1994ys}. In the present analysis, we not only obtain the total entropy as a Noether charge, but we can indeed identify $J_S^\mu= J_0^\mu+\bar{J}_0^\mu$ (with $\Lambda_0 = -2\pi$) as a local entropy current.\footnote{ This interpretation seems to make sense also for the ${\cal O}(\epsilon)$ part of the entropy current. For example, while it is obviously conserved on-shell, one can check that off-shell it satisfies a formal analog of the {\it adiabaticity equation} used in \cite{Haehl:2015pja,Haehl:2014zda}:
\begin{equation}
  \partial_\mu J^\mu_S = \frac{2\pi^2 c}{3} (\partial_\tau^3 + \partial_\tau) (\bar\partial \epsilon  + \partial\bar \epsilon) = - \beta_\mu \partial_\nu  T^{\mu\nu}_\text{h}
\end{equation}
where $\beta^\tau =2\pi$, $\beta^\sigma = 0$ and $T_{\text{h},uu}$ is defined in \eqref{eq:noetherTh}.} 
This combination of currents corresponds to equal shifts of $\epsilon$ and $\bar\epsilon$ (i.e., a diffeomorphism in the $\tau$ direction). This is again very reminiscent of hydrodynamics, where it has recently been established that (even out of equilibrium) the entropy current can be understood as a Noether current associated with an emergent gauge symmetry of thermal diffeomorphisms \cite{Haehl:2015pja,Haehl:2014zda,Haehl:2015foa}. We return to this point in the discussion section.

\vspace{10pt}
\section{Real-Time Propagators}
\label{prop}

\begin{figure}
\begin{center}
\includegraphics[width=.4\textwidth]{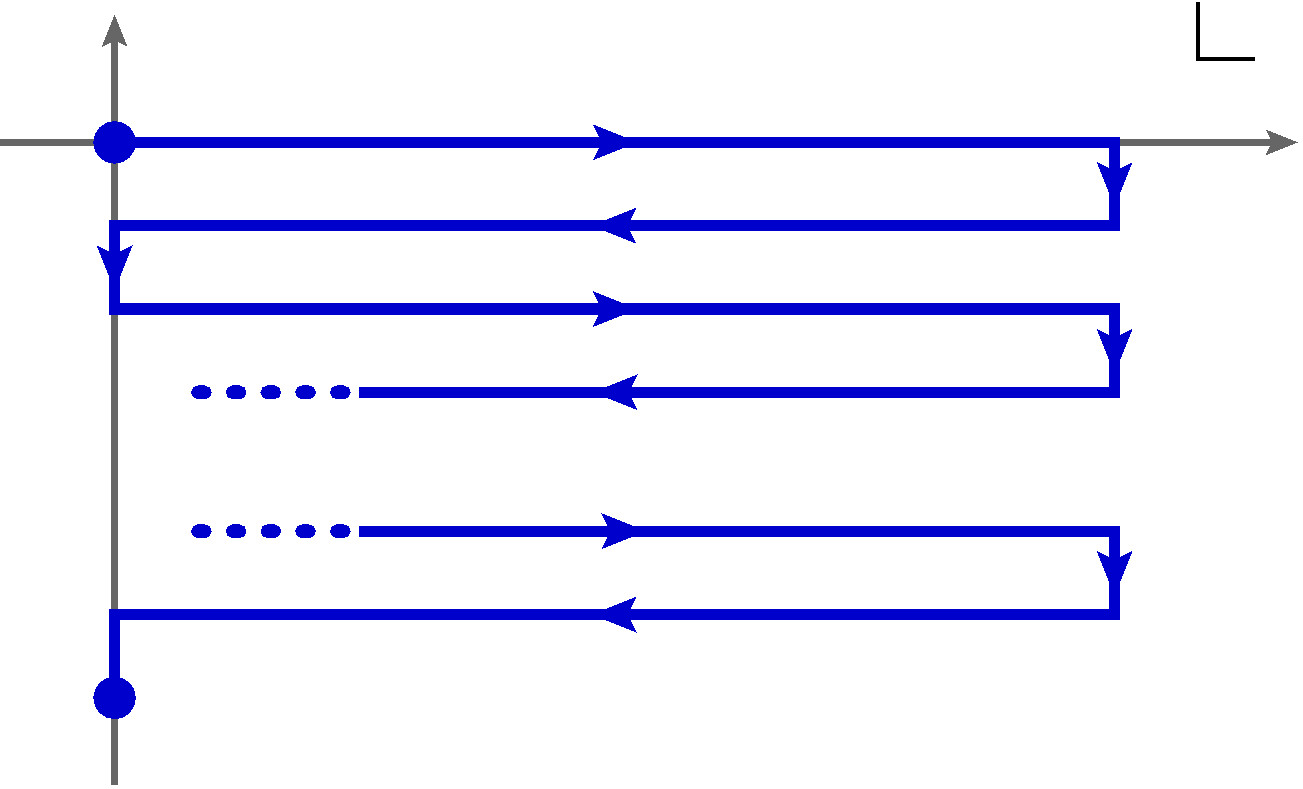}
\put(-11,102){$t$}
{\tiny
\put(-50,91){$1$}
\put(-50,79){$2$}
\put(-22,80){$-i\varphi$}
\put(-50,67){$3$}
\put(-178,69){$-i\varphi$}
\put(-50,56){$4$}
\put(-22,57){$-i\varphi$}
\put(-60,37){$2k-1$}
\put(-52,26){$2k$}
\put(-22,27){$-i\varphi$}
\put(-183,10){$-2\pi i$}
\put(-168,91){$0$}
}\normalsize
\end{center}
\caption{We choose the $k$-OTO contour in the complex time plane to be such that all legs are equally separated by $\varphi$ in the imaginary direction. We often set either $\varphi=\frac{\pi}{k}$ (equal separations) or $\varphi = \delta \ll 1$ (small separation limit). The legs of the contour are labelled by indices $\alpha,\beta,\ldots=1,\ldots,2k$.}
\label{fig:otocontour}
\end{figure}
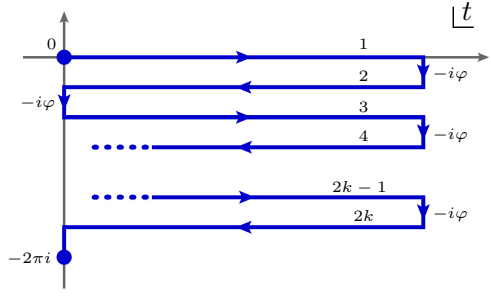

We now wish to compute the Lorentzian versions of these propagators, which we later utilize for perturbative calculations. The soft mode propagators are obtained by inverting the momentum space quadratic action and Fourier transforming to position space. There are different Lorentzian propagators, defined by different boundary conditions in time,  which are reflected in the chosen integration contour in the complex frequency plane. That is, to obtain a Lorentzian propagator we first perform the simple Wick rotation $\omega_E \mapsto i\omega$,  but in addition we need to specify a contour for avoiding poles of \eqref{eq:epsProp}. The different propagators take the form
\begin{equation}\label{eq:GijInt}
G^{ab} (t,\sigma) \equiv \langle \epsilon^a(t,\sigma)\, \epsilon^b(0,0) \rangle = -\frac{24\pi}{c} \frac{1}{(2\pi)^2} \int dk \int_{\mathcal{C}^{ab}} d\omega \, \frac{e^{-i \omega t} \, e^{i k \sigma}}{\omega(\omega^2+1)(\omega+k)} \,,
\end{equation}
where $a,b\in\{\text{av},\text{dif}\}$ specifies the average and difference combinations in the Schwinger-Keldysh doubled theory, and $\mathcal{C}^{a,b}$ is an appropriate contour in the $\omega$ plane:
\begin{equation}
\epsilon^\text{av}=\frac{\epsilon^1 + \epsilon^2}{2} \,\qquad \epsilon^\text{dif} = \epsilon^1 - \epsilon^2 \,.
\end{equation}
where the labels $\{1,2\}$ refer to the forward and backward parts of the Schwinger-Keldysh contour. The time contour appropriate for higher-OTO observables is shown in Fig.\ \ref{fig:otocontour}. The Schwinger-Keldysh case corresponds to $k=1$ in that figure.

The correlators defined by \eqref{eq:GijInt} give rise, in particular, to the retarded, advanced and Keldysh propagators:
 \begin{equation}
 G^R = -iG^{\text{av,dif}},\qquad G^A = -iG^{\text{dif,av}}, \qquad G^K = -2iG^{\text{av,av}}\,,
 \end{equation}
 while $G^{\text{dif,dif}} = 0$.

\vspace{10pt}
\paragraph{Retarded propagator:}
Indeed, to calculate the retarded propagator, we impose the following boundary conditions: $(i)$ $G^R(t< 0,\sigma)=0$ and $(ii)$ $G^R(t,|\sigma|\rightarrow \infty) =0$.
This can be ensured as follows: we take the contour of integration for the $\omega$-integral to go above all poles at $\omega\in\{-i,0,i\}$. At negative times, we close the contour in the upper half plane and don't pick up any poles. This ensures part $(i)$ of the boundary condition. 
Consistently closing the contour at infinity then implies that at late times then the propagator picks up contributions from all poles, including those in the upper half-plane. 

\begin{figure}
\begin{center}
\includegraphics[width=.8\textwidth]{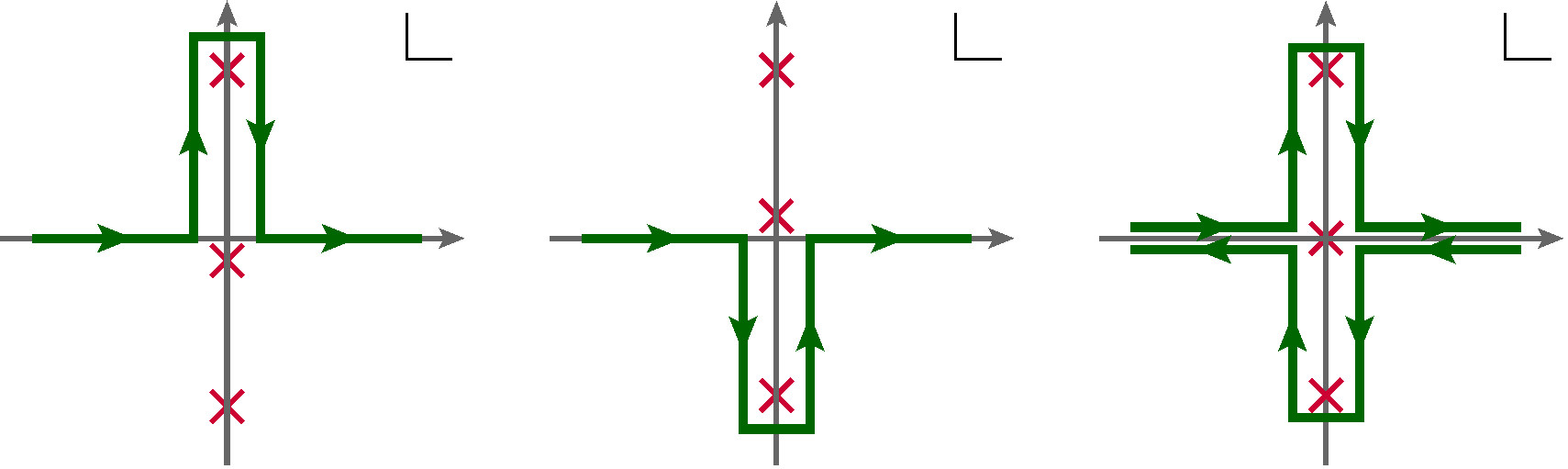}
\put(-328,35){$\mathcal{C}^R$}
\put(-203,60){$\mathcal{C}^A$}
\put(-80,60){$\mathcal{C}^K$}
\put(-12,95){$\omega$}
\put(-136,95){$\omega$}
\put(-260,95){$\omega$}
\end{center}
\caption{Contours in the complex $\omega$-plane, defining the retarded, advanced, and Keldysh (symmetric) correlation functions, respectively. The contour ${\cal C}^K = {\cal C}^R - {\cal C}^A$. The red crosses denote the poles of the integrand at $\omega \in \{-i,0\pm i \varepsilon,i\}$, where $\varepsilon>0$ is a small regulator that enforces consistent boundary conditions.}
\label{fig:contours}
\end{figure}
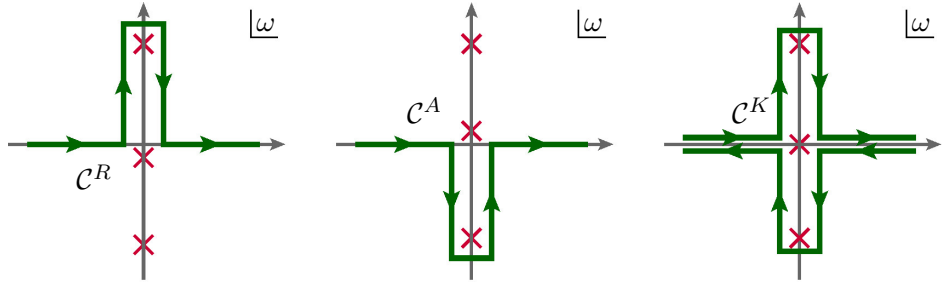

Since the subsequent $k$-integral will otherwise have an ambiguity, we should furthermore regulate the pole at $\omega=0$. We achieve this by slightly shifting this pole in the negative imaginary direction. Explicitly, we define the retarded propagator as follows:
\begin{equation}\label{eq:GijInt2}
\begin{split}
G^R(t,\sigma) &\equiv -\frac{24\pi}{c} \frac{1}{(2\pi)^2} \int dk \int_{\mathcal{C}^{R}} d\omega \, \frac{e^{-i \omega t} \, e^{i k \sigma}}{(\omega+i\varepsilon)(\omega^2+1)(\omega+k)} 
\end{split}
\end{equation}
where ${\cal C}^R$ is the leftmost contour shown in Fig.\ \ref{fig:contours} and $0<\varepsilon\ll 1$ is a regulator.
After performing the $\omega$-integral, the $k$-integral now has to be done by similar means. The integrand for the $k$-integral also has poles at $k+i\varepsilon \in\{-i,0,i\}$. Thanks to the regulator $\varepsilon$, we can simply let the $k$-contour run along the real line without having to make any further choices. We pick up poles in the lower or upper half $k$-plane, depending on the sign of $\sigma$. This gives
\begin{equation}
G^R(t,\sigma)=- \frac{12\pi}{c} \,  \Theta(t)\left[\Theta (-\sigma) \, e^{t+\sigma}-\Theta (\sigma) \, \left(e^{-(t+\sigma)} -2\,e^{-(t+\sigma)\varepsilon} \right)\, 
  \right] \,.
  \label{eq:GR}
\end{equation}
Note that time ordering for a chiral sector also implies spatial ordering, e.g. right movers can only influence events to their right in a retarded propagator. In other words: the correlator \eqref{eq:GR} allows for an exponentially growing mode in time, but it decays spatially both for $\sigma \rightarrow \infty$ as well as $\sigma \rightarrow - \infty$. 

\vspace{10pt}
\paragraph{Advanced propagator:}
The advanced correlator is computed by similar reasoning: the associated contour is shown second in Fig.\ \ref{fig:contours} and goes below all poles. The $\varepsilon$-prescription enforcing the suitable boundary condition for the pole at $\omega=0$ now shifts poles in the opposite direction, i.e., $\omega \rightarrow \omega -i\varepsilon$. Then the $k$-integral gives the correct (advanced) boundary conditions, $(i)$ $G^A(t>0,\sigma)=0$ and $(ii)$ $G^A(t,|\sigma|\rightarrow \infty) =0$. We find:
\begin{equation}
G^A(t,\sigma)= -\frac{12\pi}{c} \,  \Theta(-t)\left[\Theta (\sigma) \, e^{-(t+\sigma)}-\Theta (-\sigma) \, \left(e^{t+\sigma} -2 \, e^{(t+\sigma)\varepsilon}\right)
  \right] \,.
\end{equation}
Note that $G^A(t,\sigma) = G^R(-t,-\sigma)$.

\vspace{10pt}
\paragraph{Symmetric (Keldysh) propagator:}
This leaves us with the symmetric Keldysh Green's function. In momentum space, this correlator is given by the fluctuation-dissipation relation, $G^K(\omega,k)= \coth (\pi \omega) (G^R(\omega,k)-G^A(\omega,k))$. The associated contour is naively just ${\cal C}^K$ encircling the three poles (see Fig.\ \ref{fig:contours}). However, in order to get the right $i\varepsilon$-prescription for the pole at $\omega=0$, we compute the retarded and the advanced parts in this expression separately, using the contours ${\cal C}^R$ and ${\cal C}^A$:
\begin{equation}
\begin{split}
G^K(t,\sigma)&=  -\frac{24\pi}{c}  \frac{1}{(2\pi)^2}  \int dk \int_{\mathcal{C}^R} d\omega \,  \coth (\pi  (\omega+i\varepsilon)) \frac{e^{-i \omega t} \, e^{i k \sigma}}{(\omega+i\varepsilon)((\omega+i\varepsilon)^2+1)(\omega+k)} \\
&\quad -\frac{24\pi}{c}  \frac{1}{(2\pi)^2}  \int dk \int_{\mathcal{C}^A} d\omega \,  \coth (\pi  (\omega-i\varepsilon)) \frac{e^{-i \omega t} \, e^{i k \sigma}}{(\omega-i\varepsilon)((\omega-i\varepsilon)^2+1)(\omega+k)}
\end{split}
\end{equation}
Evaluating the contour integrals and dropping all irrelevant dependence on $\varepsilon$, we get:
{
\begin{equation}
\begin{split}
G^K(t,\sigma) &= \frac{6 i}{c}  \left[\Theta (-\sigma) \, \left(\big(  2(t+\sigma)-3 \big) 
\,e^{t+\sigma} -4 \, \Theta(-t)\,(t+\sigma) \,e^{(t+\sigma)\varepsilon} \right) \right. \\
&\qquad\;\;\;\, \left. -\Theta  (\sigma) \left(\big(2(t+\sigma)+3\big) 
\,e^{-(t+\sigma)}-4 \, \Theta(t)\,(t+\sigma) \,e^{-(t+\sigma)\varepsilon} \right)
\right] \,.
\end{split}
\end{equation}
}

This prescription is consistent with the fluctuation-dissipation theorem and the spatial boundary condition $G(t,|\sigma|\rightarrow \infty) =0$ for all propagators.

\vspace{10pt}
\paragraph{Wightman correlators:}
Once we have all the propagators in the average and difference basis, we can also find the Wightman functions, i.e., the propagators in the original contour basis. On general grounds these obey:
\begin{equation}
\label{eq:prop}
\begin{split}
iG^F(t,\sigma) &\equiv \langle {\cal T}_{SK}\, \epsilon^1(t,\sigma)\,  \epsilon^1 (0,0) \rangle = \frac{i }{2}\left(G^K +  G^R + G^A \right)(t,\sigma) \,,  \\
iG^< (t+i\varphi,\sigma)&\equiv \langle{\cal T}_{SK}\, \epsilon^1(t,\sigma)\,  \epsilon^2 (0,0) \rangle =  \frac{i }{2}\left(G^K -G^R + G^A \right) (t+i\varphi,\sigma) \,, \\
iG^>(t-i\varphi,\sigma) &\equiv  \langle{\cal T}_{SK}\, \epsilon^2(t,\sigma)\,  \epsilon^1 (0,0) \rangle = \frac{i }{2}\left(G^K +G^R - G^A \right)(t-i\varphi,\sigma)\,,  \\
iG^{\bar{F}}(t,\sigma) &\equiv \langle{\cal T}_{SK}\, \epsilon^2(t,\sigma)\,  \epsilon^2 (0,0) \rangle = \frac{i }{2}\left(G^K - G^R - G^A \right)(t,\sigma) \,.
\end{split}
\end{equation}
where ${\cal T}_{SK}$ denotes Schwinger-Keldysh time ordering along the contour and $-i\varphi$ is the imaginary time shift of the second part of the contour with respect to the first one (see Fig.\ \ref{fig:otocontour}).
As expected on general grounds, the sum $G^F - G^< - G^> + G^{\bar F} =0$. As single-time Wightman functions, we write
\begin{equation}
\label{prop2}
\begin{split}
iG^F(t,\sigma) &\equiv \langle {\cal T}\,\epsilon(t,\sigma)\epsilon(0,0) \rangle   \,,  \\
iG^<(t,\sigma)  & \equiv \langle \epsilon(0,0)\epsilon(t,\sigma) \rangle  \,, \\
iG^>(t,\sigma)  &\equiv \langle \epsilon(t,\sigma)\epsilon(0,0) \rangle  \,,  \\
iG^{\bar{F}}(t,\sigma) & \equiv \langle \overline{\cal T}\,\epsilon(t,\sigma)\epsilon(0,0) \rangle   \,,
\end{split}
\end{equation}
where ${\cal T}$  ($\overline{\cal T}$) are the usual (anti-)time ordering operations.

For explicit computations related to chaos, we often focus on the exponentially growing contribution to these Wightman functions, which can be written compactly as

\begin{equation}
\begin{split}
\langle {\cal T}_{SK}\, \epsilon^\alpha(t,\sigma)\,  \epsilon^\beta (0,0) \rangle \Big{|}_{\text{exp.\ growing}}
   & =- \frac{3}{c} \Big\{ \Theta(t) \, \Theta(-\sigma) \, \big[ 2(t+\sigma) - 3 - R_{\alpha\beta} ) \big] \, e^{t+\sigma} \, e^{i\varphi(\beta-\alpha)} \\
   &\qquad + \Theta(-t) \, \Theta(\sigma) \, \big[ 2(-t-\sigma)-3- R_{\beta\alpha}) \big] \, e^{-(t+\sigma)}\,e^{-i\varphi(\beta-\alpha)} \Big\}
\end{split}
\end{equation}
where $\alpha,\beta \in \{1,2\}$ encode which of the four Wightman functions \eqref{eq:prop} we are considering, and 
\begin{equation}
  R_{\alpha\beta} \equiv \begin{pmatrix} R_{11} & R_{12} \\ R_{21} & R_{22} \end{pmatrix}= \begin{pmatrix} -2\pi i & 2(\pi-\varphi) i \\ 2(\varphi-\pi) i & 2 \pi i \end{pmatrix} \,.
\end{equation}
Note that these expressions are particularly simple for $\varphi = \pi$, i.e., the two segments of the Schwinger-Keldysh contour at equal separation around the thermal circle. In that case, we get
\begin{equation}\label{eq:expGrowSK}
\begin{split}
\langle {\cal T}_{SK}\, \epsilon^\alpha(t,\sigma)\,  \epsilon^\beta (0,0) \rangle \Big{|}^{\varphi = \pi}_{\text{exp.\ growing}}
   & = -\frac{3}{c}\,(-)^{(\beta-\alpha)}  \Big\{ \Theta(t) \, \Theta(-\sigma) \, \big[ 2(t+\sigma+ (-)^\alpha \, \delta_{\alpha\beta}\,\pi i ) - 3 \big] \, e^{t+\sigma} \\
   &\qquad\;\; + \Theta(-t) \, \Theta(\sigma) \, \big[ 2(-t-\sigma + (-)^\alpha \, \delta_{\alpha\beta}\,\pi i ) - 3 \big] \, e^{-(t+\sigma)} \Big\}\,.
\end{split}
\end{equation}

\vspace{10pt}
\paragraph{Reproducing 1-dimensional results of the Schwarzian theory:}
As a consistency check, we can verify that our propagators reproduce those of the  well-known Schwarzian theory that describes the low-energy dynamics of the SYK model. To dimensionally reduce, we integrate over the spatial coordinate $\sigma$. This yields: 
\begin{equation}
\begin{split}
\label{eq:smear}
 \int_{-\infty}^\infty d\sigma \; G^R(t,\sigma) &= \Theta(t) \left[ -\frac{24\pi}{c} \frac{1}{\varepsilon}  + \frac{24\pi}{c} \left( t-\sinh t \right) + {\cal O}(\varepsilon) \right]\,,\\
 \int_{-\infty}^\infty d\sigma \; G^A(t,\sigma) &= \Theta(-t) \left[ -\frac{24\pi}{c} \frac{1}{\varepsilon}  - \frac{24\pi}{c} \left( t-\sinh t \right) + {\cal O}(\varepsilon) \right]\,,\\
  \int_{-\infty}^\infty d\sigma \; G^K(t,\sigma) &= \frac{24 i}{c} \frac{1}{\varepsilon^2} - \frac{24 i }{c} \left[ \frac{t^2-\pi^2}{2} -t \, \sinh t + \frac{\pi^2}{2} + \frac{5}{2} \, \cosh t  \right] + {\cal O}(\varepsilon)\,.\\
\end{split}
\end{equation}
The pieces which are divergent as $\varepsilon \rightarrow 0$ are just a constants and can be set to zero by a gauge choice. These are the pieces familiar from the analytic continuation of the N$AdS_2$ propagator (eq.\ (4.28) in \cite{Maldacena:2016upp}) with a divergence absorbed in the constant gauge mode $a$. 
The finite terms have precisely the form of the SYK or N$AdS_2$ result.

\vspace{10pt}
\paragraph{Propagators on higher-OTO contours:}
If a $\langle\epsilon\epsilon\rangle$-propagator of interest originates from a higher-OTO computation, the insertions of the soft mode occur on a contour with more switchbacks. The only difference will be a modification of the phase shift $\varphi$. Indeed, on a $k$-OTO contour where all legs of the contour are separated by $i\varphi$, the generalization of \eqref{eq:prop} reads as
\begin{equation}
 \langle {\cal T}_{SK} \; \epsilon^\alpha(t,\sigma) \, \epsilon^\beta(0,0)\rangle =
 \left\{ \begin{aligned}
 & \frac{i}{2} \left( G^K - (-)^\alpha\, G^R - (-)^\alpha\,  G^A \right) (t , \, \sigma) \qquad\quad\; (\alpha=\beta) \\
 & \frac{i}{2} \left( G^K + G^R -  G^A \right) (t+(\beta-\alpha) \; i \varphi , \, \sigma) \qquad (\alpha>\beta) \\
 & \frac{i}{2} \left( G^K - G^R +  G^A \right) (t+(\beta-\alpha) \; i \varphi , \, \sigma) \qquad (\alpha<\beta)
 \end{aligned} \right.
\end{equation}
where $\alpha,\beta = 1,\ldots,2k$ label the contour legs and as shown in Fig.\ \ref{fig:otocontour}. 
We use a convention where the labels of segments increase in the order in which they are encountered along the contour, starting at the initial state. See Fig.\ \ref{fig:otocs} for illustration.
The exponentially growing contribution, i.e., the generalization of \eqref{eq:expGrowSK}, is
\begin{equation}
\label{eq:epsepsexp}
\begin{split}
\langle {\cal T}_{SK}\, \epsilon^\alpha(t,\sigma)\,  \epsilon^\beta (0,0) \rangle \Big{|}_{\text{exp.\ growing}}
   & = -\frac{3}{c}\,  \Big\{ \Theta(t) \, \Theta(-\sigma) \, \big[ 2(t+\sigma) - 3 - R_{\alpha\beta} \big] \, e^{t+\sigma+(\beta-\alpha)\varphi i} \\
   &\qquad\;\; + \Theta(-t) \, \Theta(\sigma) \, \big[ 2(-t-\sigma ) - 3 - R_{\beta\alpha} \big] \, e^{-(t+\sigma)-(\beta-\alpha)\varphi i} \Big\}
\end{split}
\end{equation}
where now
\begin{equation}
  R_{\alpha\beta} = \left\{ \begin{aligned} & \text{sgn}(\beta-\alpha) \,2\pi i+ (\alpha-\beta)2 \varphi i \qquad (\alpha\neq \beta) \\
  & (-)^\alpha \, 2\pi i \qquad\qquad\qquad\qquad\qquad\;\; (\alpha=\beta) 
  \end{aligned} \right.
\end{equation}

\vspace{10pt}
\paragraph{Propagators for the anti-holomorphic mode $\bar\epsilon$:}
One can repeat all of the above analysis for the field $\bar\epsilon(z,\bar z)$ which breaks anti-holomorphic conformal transformations $\bar z \mapsto \bar z + \bar\epsilon(\bar z)$. Its Euclidean propagator was given in \eqref{eq:epsPropBar}.
The real time propagators for $\bar\epsilon$ with retarded, advanced, and Keldysh boundary conditions, are given by
\begin{equation}
\begin{split}
\overline{G}^R(t,\sigma) &= \frac{12\pi}{c} \,  \Theta(t)\left[\Theta (-\sigma) \, \left(e^{-(t-\sigma)} -2 \, e^{-(t-\sigma)\varepsilon}\right) 
-\Theta (\sigma) \, e^{t-\sigma}  \right] \,,\\
\overline{G}^A(t,\sigma)&= -\frac{12\pi}{c} \,  \Theta(-t)\left[\Theta (-\sigma) \, e^{-(t-\sigma)}
-\Theta (\sigma) \, \left(e^{t-\sigma} -2 \, e^{(t-\sigma)\varepsilon}\right)  \right] \,,\\
\overline{G}^K(t,\sigma) &= -\frac{6 i}{c}  \left[\Theta (-\sigma) \, \left(\big(  2(t-\sigma)+3 \big) 
\,e^{-(t-\sigma)} -4 \, \Theta(t)(t-\sigma) \,e^{-(t-\sigma)\varepsilon} \right) \right. \\
&\qquad\quad\;\;\; \left. -\Theta  (\sigma) \left(\big(2(t-\sigma)-3\big) 
\,e^{t-\sigma}-4 \, \Theta(-t)(t-\sigma) \,e^{(t-\sigma)\varepsilon} \right)
\right] \,.
\end{split}
  \label{eq:GRb}
\end{equation}

The exponentially growing parts of the propagators read
\begin{equation}
\begin{split}
\langle {\cal T}_{SK}\,\bar \epsilon^\alpha(t,\sigma)\, \bar \epsilon^\beta (0,0) \rangle \Big{|}_{\text{exp.\ growing}}
   & = -\frac{3}{c}\,  \Big\{ \Theta(t) \, \Theta(\sigma) \, \big[ 2(t-\sigma) - 3 - R_{\alpha\beta} \big] \, e^{t-\sigma+(\beta-\alpha)\varphi i} \\
   &\qquad + \Theta(-t) \, \Theta(-\sigma) \, \big[ 2(-t+\sigma ) - 3 - R_{\beta\alpha} \big] \, e^{-(t-\sigma)-(\beta-\alpha)\varphi i} \Big\}
\end{split}
\end{equation}
This is, of course, similar to \eqref{eq:epsepsexp}, but covers the spacetime quadrants where $\text{sgn}(t) = \text{sgn}(\sigma)$.

Note the following relations between various Green's functions:
\begin{equation}
\begin{split}
\overline{G}^R(t,\sigma) =G^R(t,-\sigma) = G^A(t,\sigma) = \overline{G}^A(t,-\sigma) \,,\qquad
\overline{G}^K(t,\sigma) = G^K(t,-\sigma) \,.
\end{split}
  \label{eq:relations}
\end{equation}

\vspace{10pt}
\section{Coupling to External Probes}
\label{external}

We now discuss the coupling of external probes to the soft mode. Consider a primary operator $X$ of dimension $(h,\bar h)$. The Euclidean two-point function is generated by the action
\begin{equation}\label{eq:Imatter}
  -I_{matter}^{(h,\bar h)}  = {\cal C}_X \int d^2 z_1 \, d^2z_2 \, \frac{J(z_1,\bar{z}_1) \, J(z_2,\bar{z}_2)}{(z_1-z_2)^{2h} \, (\bar{z}_1 - \bar{z}_2)^{2\bar h}} \,.
\end{equation}
Under a conformal transformation $ (z,\bar z)= (f(u), \bar{f}(\bar{u}))$, the integral transforms as:
\begin{equation}\label{eq:Imatter3}
  -I_{matter}^{(h,\bar h)}  = {\cal C}_X \int d^2 u_1  d^2u_2 \,  \left[\frac{ \partial f(u_1) \, \partial f(u_2)  }{\big(f(u_1)-f(u_2)\big)^{2}} \right]^h \, \left[\frac{ \bar\partial \bar f(\bar{u}_1) \, \bar\partial \bar{f}(\bar{u}_2) }{\big(\bar f(\bar{u}_1)-\bar f(\bar{u}_2)\big)^{2}} \right]^{\bar h} \,J(u_1,\bar{u}_1) \, J(u_2,\bar{u}_2) \,,
\end{equation}
which is invariant for $f$ and $\bar {f}$ being $SL(2,\mathbb{R})$ transformations. 

Now consider the map to the thermal state and the symmetry breaking modes: $(z,\bar z) = \big(e^{-i(u+\varepsilon(u,\bar u))},\, e^{i(\bar u+\bar \varepsilon(u,\bar u))} \big)$. This gives an action
\begin{equation}
\begin{split}
-I_{matter}^{(h,\bar h)} 
 &= \frac{{\cal C}_X}{2^{2(h+\bar h)}} \int d^2u_1\, d^2u_2 \, \frac{J(u_1,\bar{u}_1)\, J(u_2,\bar{u}_2)}{\sin^{2h}\!\left( \frac{u_1-u_2}{2} \right) \, \sin^{2\bar h}\!\left( \frac{\bar{u}_1-\bar{u}_2}{2} \right)} \, \sum_{p\geq 0}  \mathcal{B}^{(p)}_{(h,\bar h)}(u_1,u_2,\bar{u}_1,\bar{u}_2) 
\end{split}
\end{equation}
where $p$ labels the order in $\epsilon$ and $\bar \epsilon$, and we defined the bilocal ``vertices'' 
\begin{equation}
\label{eq:verticesEucl}
\begin{split}
\mathcal{B}^{(0)}_{(h,\bar h)} &= 1 \,,\\
\mathcal{B}^{(1)}_{(h,\bar h)} &=  
h \left( \partial \epsilon_1 +\partial \epsilon_2  - \frac{\epsilon_1 - \epsilon_2}{\tan \left( \frac{u_1-u_2}{2} \right)}\right) 
+ 
\bar h \left(  \bar{\partial} \bar{\epsilon}_1+ \bar{\partial} \bar{\epsilon}_2   - \frac{\bar\epsilon_1 - \bar\epsilon_2}{\tan \left( \frac{\bar{u}_1-\bar{u}_2}{2} \right)} \right)
\,,\\
 %
  \mathcal{B}^{(2)}_{(h,\bar h)} &=
 \frac{1+h+h\,\cos {u}_{12}}{4\, \sin^2\left(\frac{{u}_{12}}{2}\right)}h\, (\epsilon_1-\epsilon_2)^2  + \frac{1+\bar{h}+\bar{h}\,\cos \bar{u}_{12}}{4\, \sin^2\left(\frac{\bar{u}_{12}}{2}\right)}\bar{h}\, (\bar\epsilon_1-\bar\epsilon_2)^2 + \frac{h\bar{h}\,(\epsilon_1-\epsilon_2)(\bar\epsilon_1-\bar\epsilon_2)}{\tan\left(\frac{u_{12}}{2}\right) \, \tan \left(\frac{\bar{u}_{12}}{2}\right)} \\
  &\qquad - \left( h\frac{\epsilon_1-\epsilon_2}{\tan\left(\frac{u_{12}}{2}\right)} +\bar{h} \frac{\bar\epsilon_1-\bar\epsilon_2}{\tan \left(\frac{\bar{u}_{12}}{2}\right)} \right) (h\,\partial\epsilon_1+\bar{h}\,\bar\partial\bar\epsilon_1+h\,\partial\epsilon_2+\bar{h}\,\bar\partial\bar\epsilon_2)\\
  &\qquad +(h\,\partial\epsilon_1+\bar{h}\,\bar\partial \bar\epsilon_1) (h\,\partial\epsilon_2+\bar{h}\,\bar\partial\bar\epsilon_2)  
  + h\bar{h}\, (\partial\epsilon_1 \, \bar\partial\bar\epsilon_1 + \partial\epsilon_2 \, \bar\partial\bar\epsilon_2 ) \\
  &\qquad + \frac{h(h-1)}{2} \big( (\partial \epsilon_1)^2 + (\partial \epsilon_2)^2 \big) + \frac{\bar{h}(\bar{h}-1)}{2} \big( (\bar\partial \bar\epsilon_1)^2 + (\bar\partial \bar\epsilon_2)^2 \big)  \,,
\end{split}
\end{equation}
and so on.
The vertex $\mathcal{B}^{(1)}_{(h,\bar h)}$ is $SL(2,\mathbb{R})$ symmetric in the sense that it is invariant under $\epsilon_j \mapsto \epsilon_j+a+ b e^{i u} + c e^{-i u}$ and similarly for $\bar\epsilon_j$. This symmetry descends from the exact $SL(2,\mathbb{R})$ symmetry of the matter action \eqref{eq:Imatter}, but takes this simple form only to leading order in $\epsilon$. 

We can make the $SL(2,\mathbb{R})$ symmetry more manifest, for example by writing $\mathcal{B}^{(1)}_{(h,\bar h)}$  as 
{\small
\begin{equation}
\begin{split}
\mathcal{B}^{(1)}_{(h,\bar h)} & =  \frac{-i}{\langle X_1 X_2 \rangle}\, \Big\{ L_0^{(1)} \left[ \epsilon_1\langle X_1 X_2 \rangle \right]  + L_0^{(2)} \left[\epsilon_2\langle X_1 X_2 \rangle\right]  - \bar{L}_0^{(1)} \left[ \bar\epsilon_1 \langle X_1X_2 \rangle \right]   - \bar{L}_0^{(2)} \left[ \bar\epsilon_2 \langle X_1 X_2 \rangle \right]  \Big\}
\end{split}
\end{equation}
}\normalsize
where $\epsilon_i = \epsilon(u_i,\bar{u}_i)$, the Euclidean two-point function $\langle X_1 X_2 \rangle \propto \sin^{-2h}\!\left( \frac{u_{12}}{2} \right) \, \sin^{-2\bar h}\!\left( \frac{\bar{u}_{12}}{2} \right)$ and superscripts on the $L_n^{(i)}$ operators indicate which pair $(u_i,\bar{u}_i)$ they act on. The action of $L_n$ generators is reviewed in Appendix \ref{sl2}.

\vspace{10pt}
\paragraph*{Lorentzian couplings:}
We now analytically continue these Euclidean results to Lorentzian signature and to potentially multi-segment contours. 
Setting $u = \tau + i \sigma \rightarrow i (t + \sigma)$ we find
\begin{equation}
\begin{split}
I_{matter}^{(h,\bar h)} 
 &= \sum_{\alpha,\beta=1}^{2k} \frac{4\,{\cal C}_X}{(2i)^{2(h+\bar h)}} \int dt_1d\sigma_1dt_2d\sigma_2 \, \frac{ \sum_{p\geq 0}  \mathcal{B}^{(p)\alpha\beta}_{(h,\bar h)}(t_1,\sigma_1,t_2,\sigma_2) \;J^\alpha(t_1,\sigma_1)\, J^\beta(t_2,\sigma_2)}{\sinh^{2h}\!\left( \frac{t_{12}-i(\alpha-\beta)\varphi+\sigma_{12}}{2} \right) \, \sinh^{2\bar h}\!\left( \frac{t_{12}-i(\alpha-\beta)\varphi-\sigma_{12}}{2} \right)} 
\end{split}
\end{equation}
where $\alpha,\beta=1,\ldots,2k$ label the segments of the $k$-OTO time contour, and we have
\begin{equation}
\label{eq:B01lorentz}
\begin{split}
\mathcal{B}^{(0)\alpha\beta}_{(h,\bar h)} (t_1,\sigma_1,t_2,\sigma_2) &= 1 \,,\\
\mathcal{B}^{(1)\alpha\beta}_{(h,\bar h)} (t_1,\sigma_1,t_2,\sigma_2) &=  -ih \left( \frac{1}{2}(\partial_{t_1}+\partial_{\sigma_1}) \epsilon_1^\alpha + \frac{1}{2}(\partial_{t_2}+\partial_{\sigma_2}) \epsilon_2^\beta  - \frac{\epsilon_1^\alpha - \epsilon_2^\beta}{\tanh \left( \frac{t_{12}-i(\alpha-\beta)\varphi+\sigma_{12}}{2} \right)} \right) \\
&\quad -i\bar h 
 \left(\frac{1}{2}(\partial_{t_1}-\partial_{\sigma_1}) \bar{\epsilon}_1^\alpha+ \frac{1}{2}(\partial_{t_2}-\partial_{\sigma_2}) \bar{\epsilon}_2^\beta - \frac{\bar\epsilon_1^\alpha - \bar\epsilon_2^\beta}{\tanh \left( \frac{t_{12}-i(\alpha-\beta)\varphi-\sigma_{12}}{2} \right)} \right) \,.
\end{split}
\end{equation}
Analogously, $\mathcal{B}^{(2)\alpha\beta}_{(h,\bar{h})}$ also reads as in \eqref{eq:verticesEucl} with the replacements $u \rightarrow i(t+\sigma)$ and $\bar{u}\rightarrow i(t-\sigma)$, and contour labels attached to the soft modes.

\vspace{10pt}
\section{Correlation Functions}
\label{correlation}

In this section we use the ingredients obtained above to perform a perturbative calculation of the OTOCs capturing the early-time Lyapunov behaviour. We discuss both the 4-point function as well as its ``maximally braided'' generalizations introduced in \cite{Haehl:2017pak}.

\vspace{10pt}
\subsection{Out-of-Time-Order 4-Point Function}

Consider the Lorentzian OTOC 4-point function with contour separation $\varphi$ in the imaginary direction:
\begin{equation}
\label{eq:4ptDef}
C_4(t,\sigma)= \frac{\langle X(t-3i\varphi ,\sigma)Y(-2i\varphi,0)X(t-i\varphi,\sigma)Y(0,0) \rangle}{\langle XX \rangle \langle YY \rangle}\,.
\end{equation}
In order to represent this correlator on a $2$-OTO time contour (with $2$ forward and $2$ backward legs), we consider $4$ copies of each operator, $X^\alpha$, labelled by a superscript $\alpha=1,\ldots,4$ indicating which segment of the contour the operator is inserted on. 

\begin{figure}
\begin{center}
\includegraphics[width=.8\textwidth]{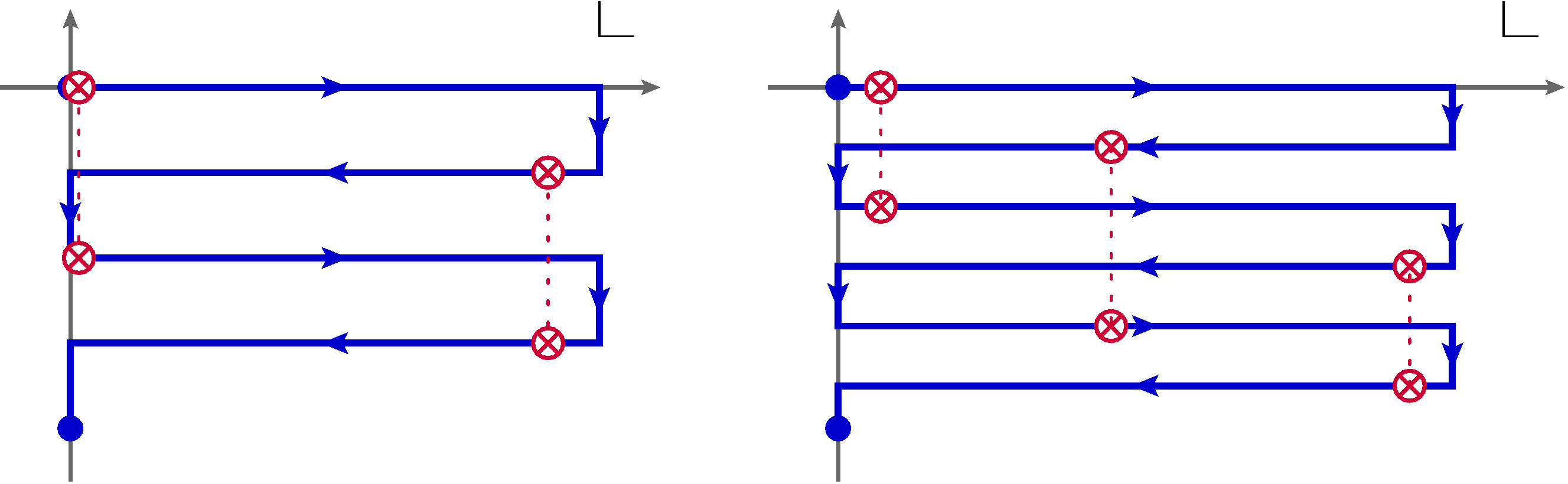}
\put(-334,95){$Y(0)$}
\put(-334,56){$Y(0)$}
\put(-255,76){$X(t)$}
\put(-255,36){$X(t)$}
\put(-151,95){{\scriptsize$Z(t_3)$}}
\put(-151,66){{\scriptsize$Z(t_3)$}}
\put(-128,80){{\scriptsize$Y(t_2)$}}
\put(-128,40){{\scriptsize$Y(t_2)$}}
\put(-62,53){{\scriptsize$X(t_1)$}}
\put(-62,26){{\scriptsize$X(t_1)$}}
\put(-216,103){$t$}
\put(-12,103){$t$}
\end{center}
\caption{Illustration of the 4- and 6-point out-of-time-ordered correlators that we consider.}
\label{fig:otocs}
\end{figure}
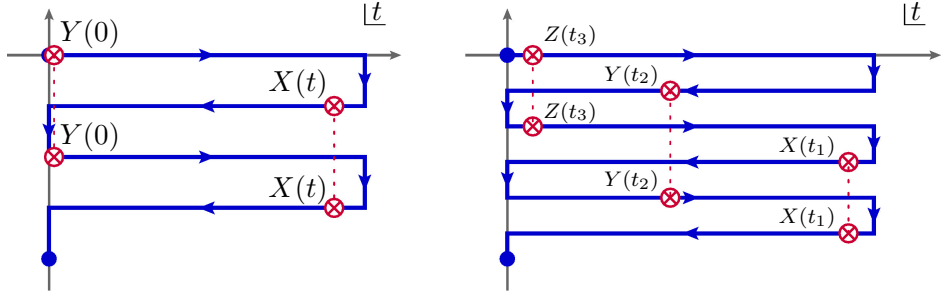

With this convention, we can represent $C_4(t,\sigma)$ on a $2$-OTO contour as
\begin{equation}
\label{eq:otoc1}
\begin{split}
C_4(t,\sigma) &= \frac{\langle {\cal T}_{{\cal C}_2} \,X^4(t,\sigma)X^2(t,\sigma)Y^3(0,0)Y^1(0,0)  \rangle}{\langle XX \rangle \langle YY \rangle}  \\
&= \Big\langle {\cal T}_{{\cal C}_2} \,\mathcal{B}^{(1)\,42}_{(h_X,\bar h_X)}(t,\sigma,t,\sigma) \; \mathcal{B}^{(1)\,31}_{(h_Y,\bar h_Y)}(0,0,0,0)\Big\rangle + {\cal O}(1/c^2 )\,,
\end{split}
\end{equation}
where ${\cal T}_{{\cal C}_2}$ denotes contour ordering along the appropriate $2$-OTO time contour and $\mathcal{B}^{(1)\,\alpha\beta}_{(h,\bar h)}$ can be read off of \eqref{eq:B01lorentz}. At lowest non-trivial order in the soft modes (i.e., in $\frac{1}{c}$ perturbation theory), this correlator decomposes into $\langle \epsilon^\alpha \epsilon^\beta \rangle$ two-point functions.
Each one of them can be represented on a 1-OTO Schwinger-Keldysh contour (or -- redundantly -- on the original 2-OTO contour). Depending on the labels $\alpha,\beta$, each of them is then one of the four Wightman correlators written down in \eqref{eq:prop}. 

We can extract the exponentially growing part of these soft mode two-point propagators from \eqref{eq:epsepsexp} and find that they contribute to the four-point OTOC as 
\begin{equation}
\label{eq:F4phi}
\begin{split}
 C_4(t,\sigma)\big{|}_\text{exp.\ growing} &= -\frac{12\pi i}{c} \, \frac{1}{\sin(\varphi)^2} \, \bigg\{ h_X h_Y\, \left[\Theta(t)\Theta(-\sigma) \, e^{t+\sigma-i\varphi} - \Theta(-t) \Theta(\sigma) \, e^{-t-\sigma+i\varphi} \right] \\
 &\qquad\qquad \qquad\quad\;\;
 +\bar h_X \bar{h}_Y\, \left[\Theta(t)\Theta(\sigma) \, e^{t-\sigma-i\varphi} - \Theta(-t) \Theta(-\sigma) \, e^{-t+\sigma+i\varphi} \right] \bigg\} 
\end{split}
\end{equation}
For the choice $\varphi = \frac{\pi}{2}$, such that the legs of the contour are equally separated around the Euclidean time circle, this result becomes
\begin{equation}
\boxed{
\begin{split}
 C_4(t,\sigma)\big{|}_\text{exp.\ growing}^{\varphi=\pi/2} &= - \frac{12\pi}{c} \, \bigg\{ h_X h_Y\, \left[\Theta(t)\Theta(-\sigma) \, e^{t+\sigma} + \Theta(-t) \Theta(\sigma) \, e^{-t-\sigma} \right] \\
 &\qquad\quad\;\;\;
 +\bar h_X \bar{h}_Y\, \left[\Theta(t)\Theta(\sigma) \, e^{t-\sigma} + \Theta(-t) \Theta(-\sigma) \, e^{-t+\sigma} \right] \bigg\} 
\end{split}
}
\end{equation}
As expected, this OTO 4-point function grows exponentially in time, but decays exponentially in space. It receives complementary contributions from the holomorphic and anti-holomorphic sectors. Moreover, we can read off the Lyapunov exponent $\lambda_L = 1 \equiv \frac{2\pi}{\beta} $ and the butterfly velocity $v_B = 1$. The {\it scrambling time} when the OTOC reaches an ${\cal O}(c^0)$ value is $t_* \equiv \log c$.\footnote{ More precisely, the scrambling time depends on ratios of $c$ and the operator dimensions $h$, $\bar{h}$, if the latter are not ${\cal O}(1)$.} This agrees with the values read off the pole-skipping location.

For comparison, the time-ordered 4-point function can be written as
\begin{equation}
\label{eq:timeordered}
\begin{split}
\frac{\langle X(t-\frac{3\pi i}{2},\sigma)X(t-\pi i,\sigma)Y(-\frac{\pi i}{2},0)Y(0,0) \rangle}{\langle XX \rangle \langle YY \rangle}
&= \Big\langle {\cal T}_{{\cal C}_2} \, \mathcal{B}^{(1)\,43}_{(h_X,\bar h_X)}(t,\sigma,t,\sigma) \; \mathcal{B}^{(1)\,21}_{(h_Y,\bar h_Y)}(0,0,0,0) \Big\rangle 
\end{split}
\end{equation}
The exponentially growing part of \eqref{eq:timeordered} vanishes, as expected for a time-ordered correlator. 

Before turning to higher-point functions, let us briefly discuss two other methods of regularizing the real-time correlation functions of interest.

\vspace{10pt}
\subsubsection*{$\delta$-Regularization: small contour separation}

A convenient tool for distinguishing the 4-point OTOC from the TOC is to set $\varphi \equiv \delta$ and expand in $\delta \ll 1$. One can check that the exponentially growing terms are  the most divergent as $\delta \rightarrow 0$. For example, we find for the leading divergence of the 4-point OTOC (now writing even the terms which are not exponentially growing in time):
\begin{equation}
\begin{split}
& \frac{\langle X(t-3i\delta,\sigma)\,Y(-2i\delta ,0)\,X(t-i\delta,\sigma)\,Y(0,0) \rangle}{\langle XX \rangle \langle YY \rangle} \\
&\qquad =  -\frac{12\pi i}{c}\frac{1}{\delta^2} \,\bigg\{ \Theta(t)\Theta(\sigma) \left[ \bar{h}_X\bar{h}_Y  e^{t-\sigma}  - {h}_X{h}_Y \left( e^{-(t+\sigma)} -  2e^{-(t+\sigma)\varepsilon} \right) \right] \\
&\qquad\qquad\qquad\quad +  \Theta(t)\Theta(-\sigma) \left[ {h}_X {h}_Y  e^{t+\sigma}  - \bar{h}_X\bar{h}_Y \left( e^{-(t-\sigma)} -  2e^{-(t-\sigma)\varepsilon} \right) \right] \\
&\qquad\qquad\qquad\quad -  \Theta(-t)\Theta(\sigma) \left[ {h}_X {h}_Y  e^{-t-\sigma}  - \bar{h}_X\bar{h}_Y \left( e^{t-\sigma} -  2e^{(t-\sigma)\varepsilon} \right) \right] \\
&\qquad\qquad\qquad\quad -  \Theta(-t)\Theta(-\sigma) \left[\bar{h}_X \bar{h}_Y  e^{-t+\sigma}  - \bar{h}_X\bar{h}_Y \left( e^{t+\sigma} -  2e^{(t+\sigma)\varepsilon} \right) \right]  \bigg\}
\end{split}
\end{equation}
In each square bracket, the first term describes the exponentially growing mode and is consistent with expanding \eqref{eq:F4phi} for small $\varphi$.\footnote{  Note that these results are also consistent with those obtained in \cite{Roberts:2014ifa}: in our notation, their case corresponds to $t> -\sigma > 0$. The regime of validity of our approach is larger, since we do not have to assume anything about the operator dimensions, apart from $h_X,h_Y,\bar{h}_X,\bar{h}_Y \ll c$.} The latter terms in the square brackets are decaying and are uninteresting for the purpose of diagnosing chaos. 

For comparison, the time-ordered 4-point function \eqref{eq:timeordered} with infinitesimal contour separation is again zero as $\delta \rightarrow 0$:
\begin{equation}
\begin{split}
& \frac{\langle X(t-3i\delta,\sigma)\,X(t-2i\delta ,0)\,Y(-i\delta,\sigma)\,Y(0,0) \rangle}{\langle XX \rangle \langle YY \rangle} \rightarrow 0 \qquad \text{for } \delta \rightarrow 0\,.
\end{split}
\end{equation}

Hence this limit serves to distinguish the two cases and allows us to focus on the chaos regime. This is useful in separating the maximally growing mode in the next subsection.

\vspace{10pt}
\subsubsection*{Real time regularization}

Another way of regularizing the correlators is by separating the real time insertion points and having the various segments of the complex time contour exactly on top of each other (i.e., $\varphi = 0$). This was used in \cite{Maldacena:2016upp}. It is straightforward to reproduce the analytic continuations of their Euclidean results for the TOC (their eq.\ (4.30)) and the OTOC (their eq.\ (4.32)) by computing the corresponding 4-point functions directly using our Lorentzian methods and the dimensionally reduced Schwinger-Keldysh propagators \eqref{eq:smear} appropriate for the Schwarzian quantum mechanics. As an illustration, let us compute the difference of the OTOC and the TOC:
\begin{equation}
\label{eq:F4def}
\begin{split}
F_4(t_1,\ldots,t_4;\sigma_1,\ldots,\sigma_4) &\equiv \frac{\big{\langle} X(t_1,\sigma_1)\rho^{-\frac{\varphi}{2\pi}}\,[Y(t_3,\sigma_3)\rho^{-\frac{\varphi}{2\pi}}\,,X(t_2,\sigma_2)\rho^{-\frac{\varphi}{2\pi}}\,] Y(t_4,\sigma_4)\rho^{-\frac{\varphi}{2\pi}}\, \rho^{\frac{4\varphi}{2\pi}}  \big{\rangle}}{\langle XX \rangle \langle YY \rangle}  \,,
\end{split}
\end{equation}
where $\rho = e^{-2\pi H}$ is the thermal density matrix defining the initial state, which evolves operators in Euclidean time according to $X(t,\sigma) \rho^\alpha = \rho^\alpha X(t+2\pi  i\alpha,\sigma)$. The fractional powers of $\rho$ serve to implement the same contour separations for the two parts of the commutator as in \eqref{eq:4ptDef} and \eqref{eq:timeordered}.

Thanks to the commutator in \eqref{eq:F4def}, there are many cancellations and we can write the full time and space dependence for $\varphi =0$ without too much effort (though we do have in mind $t_1 \approx t_2$ and $t_3 \approx t_4$). The result reads as follows: 
\begin{equation}
\begin{split}
  F_4 &= -\frac{12\pi i}{c} \bigg\{ 
  2 h_X h_Y \left[\Theta(t_{23}) \Theta(\sigma_{23})-\Theta(-t_{23}) \Theta(-\sigma_{23})\right] \, \coth\left(\frac{t_{12}+\sigma_{12}}{2}\right)\coth\left(\frac{t_{34}+\sigma_{34}}{2}\right)\\
  &\qquad\qquad + 2 \bar{h}_X \bar{h}_Y \left[\Theta(t_{23}) \Theta(-\sigma_{23})-\Theta(-t_{23}) \Theta(\sigma_{23})\right] \, \coth\left(\frac{t_{12}-\sigma_{12}}{2}\right)\coth\left(\frac{t_{34}-\sigma_{34}}{2}\right)\\
  &\qquad + \frac{h_X h_Y}{\sinh\left(\frac{t_{12}+\sigma_{12}}{2}\right)\sinh\left(\frac{t_{34}+\sigma_{34}}{2}\right)} \left[\Theta(-\sigma_{23}) \, e^{\frac{t_{14}+t_{23}+\sigma_{14}+\sigma_{23}}{2}}-\Theta(\sigma_{23}) \, e^{\frac{-t_{14}-t_{23}-\sigma_{14}-\sigma_{23}}{2}}\right]\\
  &\qquad + \frac{\bar{h}_X \bar{h}_Y}{\sinh\left(\frac{t_{12}-\sigma_{12}}{2}\right)\sinh\left(\frac{t_{34}-\sigma_{34}}{2}\right)} \left[\Theta(\sigma_{23}) \, e^{\frac{t_{14}+t_{23}-\sigma_{14}-\sigma_{23}}{2}}-\Theta(-\sigma_{23}) \, e^{\frac{-t_{14}-t_{23}+\sigma_{14}+\sigma_{23}}{2}}\right] \bigg\}
\end{split}
\end{equation}
Dependence on $\sigma_{ij}$ is again never exponentially growing, while dependence on $t_{ij}$ admits exponential growth. The chaos regime can be explored, e.g., by setting $\sigma_{1,\ldots,4}= 0$, $t_1 \sim t_2 \sim t$, and $t_3 \sim t_4 \sim 0$, in which case the correlators grow as $F_4 \sim e^{|t|-t_*}$ with scrambling time $t_* \equiv \log c$.

\vspace{10pt}
\subsection{Out-of-Time-Order 6-point Function}

The space of all possible $n$-point correlation functions (most of which are to out-of-time-order, albeit to different extent) provides a large set of quantum field theory observables \cite{Haehl:2017qfl,Haehl:2017eob,Halpern:2017abm}. It would be desirable to understand the physics encoded by all of these. While we shall not attempt to understand the subtleties of arbitrary higher-point OTOCs, there exists a particular generalization of the 4-point function studied above, which diagnoses more fine-grained features of quantum chaos in a useful and simple way \cite{Haehl:2017pak,Qi:2018rqm,Lam:2018pvp}. In this section we study the 6-point version of this observable, and subsequently proceed with the general case.

Consider the Lorentzian 6-point function
{\small
\begin{equation}
\begin{split}
&F_6(t_1,t_2,t_3;\sigma_1,\sigma_2,\sigma_3)  \\
&\equiv \frac{\big{\langle} X(t_1,\sigma_1) \rho^{-\frac{\varphi}{2\pi}} \, \big[Y(t_2,\sigma_2)\rho^{-\frac{\varphi}{2\pi}},X(t_1,\sigma_1)\rho^{-\frac{\varphi}{2\pi}}\,\big]\, \big[  Z(t_3,\sigma_3) \rho^{-\frac{\varphi}{2\pi}},Y(t_2,\sigma_2) \rho^{-\frac{\varphi}{2\pi}}\,\big]\,Z(t_3,\sigma_3) \rho^{-\frac{\varphi}{2\pi}} \, \rho^{\frac{6\varphi}{2\pi}} \big{\rangle}}{\langle XX \rangle \langle YY \rangle \langle ZZ \rangle}
\end{split}
\label{eq:F6def}
\end{equation}
}%
The combination \eqref{eq:F6def} generalizes \eqref{eq:F4def} and was identified in \cite{Haehl:2017pak} as a particularly good candidate for a higher-point OTOC that diagnoses quantum chaos in a maximally fine-grained way. The basic object of interest in $F_6$ is the term where all the commutators are dropped and operators occur in the order as indicated. This is also the object depicted in Fig.\ \ref{fig:otocs}. The commutators in \eqref{eq:F6def} are simply for convenience: they serve to subtract off all pieces which are less out-of-time-order and have a slower exponential growth. They allow us to focus on just the piece of the 6-point function that is responsible for the latest-time signature of scrambling. 

We refer to $F_6$ (or more precisely: the term in $F_6$ obtained by removing all commutator brackets) as the 6-point function which is both {\it maximally out-of-time-order} and {\it maximally braided} for the following reasons:
\begin{itemize}
\item {\it Maximally out-of-time-order} means that it cannot be represented on a time contour with less than 3 forward and 3 backward legs. This is the maximum number of switchbacks in time required to represent any 6-point function. $F_6$ being maximally OTO is a feature of the Lorentzian times (in Fig.\ \ref{fig:otocs} it requires that either $t_1>t_2>t_3$, or $t_3>t_2>t_1$).
\item {\it Maximally braided} refers to the ordering of the Euclidean times, i.e., the imaginary times of the insertions in Fig.\ \ref{fig:otocs}. Projecting the insertions in Fig.\ \ref{fig:otocs} onto the Euclidean circle (imaginary axis) and connecting equal operators by lines representing propagators, one obtains a picture of propagators mutually braided in Euclidean time. This time ordering leads to the longest possible exponential growth for a 6-point function.
\end{itemize}

To study the essential signatures of quantum chaos, it is again convenient to extract the exponentially growing features by setting $\varphi \equiv \delta$ and focus on the leading term in the small-$\delta$ expansion. For simplicity, we will assume that the operators $X$, $Y$, $Z$ all have the same dimensions $(h,\bar h)$ (with $h,\bar h \ll c$).
For the 6-point function defined above, we find the following exponentially growing contribution: 
{\small
\begin{equation}
\begin{split}
& F_6  (t_1,t_2,t_3;\sigma_1,\sigma_2,\sigma_3)\big{|}_\text{exp. growing} \\
 &\quad= \frac{32\pi^2}{c^2}\frac{1}{\delta^4} \bigg\{ -\Theta(t_{12})\Theta(t_{23}) \, e^{t_{13}} \, \Big[ h^3(1+2h) \, \Theta(\sigma_{21})\Theta(\sigma_{32})\,e^{-\sigma_{31}} + \bar{h}^3(1+2\bar{h}) \, \Theta(\sigma_{12})\Theta(\sigma_{23})\,e^{-\sigma_{13}}\\
 & \qquad\qquad\qquad\qquad\qquad\qquad\qquad\qquad +2 h^2\bar{h}^2 \left( \Theta(\sigma_{12})\Theta(\sigma_{32}) \, e^{-\sigma_{12}-\sigma_{32}} +  \Theta(\sigma_{21})\Theta(\sigma_{23}) \, e^{-\sigma_{21}-\sigma_{23}} \right)\Big]\\
&\qquad\qquad\qquad\;\; - \Theta(t_{32})\Theta(t_{21}) \, e^{t_{31}} \, \Big[ h^3(1+2h) \, \Theta(\sigma_{12})\Theta(\sigma_{23})\,e^{-\sigma_{13}} + \bar{h}^3(1+2\bar{h}) \, \Theta(\sigma_{21})\Theta(\sigma_{32})\,e^{-\sigma_{31}}\\
 & \qquad\qquad\qquad\qquad\qquad\qquad\qquad\qquad + 2 h^2\bar{h}^2 \left( \Theta(\sigma_{12})\Theta(\sigma_{32}) \, e^{-\sigma_{12}-\sigma_{32}} +  \Theta(\sigma_{21})\Theta(\sigma_{23}) \, e^{-\sigma_{21}-\sigma_{23}} \right)\Big]\\
&\qquad \; +  \Theta(t_{12})\Theta(t_{32}) \, e^{t_{12}+t_{32}} \Big[ h^3(1+2h) \Theta(\sigma_{21}) \Theta(\sigma_{23}) \, e^{-\sigma_{21}-\sigma_{23}} + \bar h^3 (1+2\bar h) \Theta(\sigma_{12})\Theta(\sigma_{32}) \, e^{-\sigma_{12}-\sigma_{32}} \\ 
 & \qquad\qquad\qquad\qquad\qquad\qquad\quad + 2 h^2\bar{h}^2 \left( \Theta(\sigma_{12})\Theta(\sigma_{23}) \, e^{-\sigma_{13}} +  \Theta(\sigma_{21})\Theta(\sigma_{32}) \, e^{-\sigma_{31}} \right) \Big]\\
&\qquad \; +  \Theta(t_{21})\Theta(t_{23}) \, e^{t_{21}+t_{23}} \Big[ h^3(1+2h) \Theta(\sigma_{12}) \Theta(\sigma_{32}) \, e^{-\sigma_{12}-\sigma_{32}} + \bar h^3 (1+2\bar h) \Theta(\sigma_{21})\Theta(\sigma_{23}) \, e^{-\sigma_{21}-\sigma_{23}} \\ 
 & \qquad\qquad\qquad\qquad\qquad\qquad\quad + 2 h^2\bar{h}^2 \left( \Theta(\sigma_{12})\Theta(\sigma_{23}) \, e^{-\sigma_{13}} +  \Theta(\sigma_{21})\Theta(\sigma_{32}) \, e^{-\sigma_{31}} \right) \Big] \bigg\}
 +{\cal O}(\delta^{-3})
\end{split} 
\label{eq:6ptres}
\end{equation}
}\normalsize
This is the expression generalizing the fine-grained chaos of \cite{Haehl:2017pak} to two-dimensional chaotic CFTs. Let us discuss a few salient features of this result: 
\begin{itemize}
 \item As in the case of the 4-point OTOC, all terms are such that they grow exponentially in time, but exponentially decay in space. 
 \item The first four lines grow exponentially in only a single time scale $|t_{13}|$. Keeping in mind the arrangement of operators of Fig.\ \ref{fig:otocs}, this timescale can be thought of as the total duration of the ``experiment''. The Lyapunov exponent is therefore $\lambda_L = 1 \equiv \frac{2\pi}{\beta}$, i.e., the same as for the 4-point OTOC. However, since the observable $F_6$ is suppressed by $\frac{1}{c^2}$, the associated $3$-OTO scrambling time is 
 \begin{equation}
    t_*^{(3)} = 2 \, t_* \equiv 2\times \frac{\beta}{2\pi} \, \log c \,,
 \end{equation}
 where $t_* = \frac{\beta}{2\pi}\log c$ is the standard 4-point OTOC scrambling time. Only after $|t_{13}| \sim t_*^{(3)}$ will the 6-point OTOC approach to an ${\cal O}(c^0)$ value. It is this non-trivial interplay between an unchanged Lyapunov exponent on the one hand, and an (obvious) higher suppression in $\frac{1}{c}$ on the other hand, that has been identified in \cite{Haehl:2017pak} as the reason for a longer characteristic scrambling time associated with higher-point OTOCs.
 \item The last four lines of \eqref{eq:6ptres} are different from the first four: they depend on two timescales, $|t_{12}|$ and $|t_{32}|$. By drawing the associated contour representation, one can easily see that in these cases the 6-point function is actually only $2$-OTO, i.e., it can be represented on a contour with only four (instead of six) legs. This is reflected in the fact that the characteristic time scale of these configurations is only $t_*$, not $2t_*$ as above: once both $|t_{12}|$ and $|t_{32}|$ become of order $t_*$, then they already outweigh the $\frac{1}{c^2}$ suppression. 

\end{itemize}

Finally, note the intricate interplay between holomorphic and anti-holomorphic conformal weights and coordinate dependence in \eqref{eq:6ptres}. This is a unique feature of the higher-dimensional case that lends much more structure to the OTOC than in a quantum mechanical system.

\vspace{10pt}
\subsection{Out-of-Time-Order $2k$-point Function}

It is not hard to generalize these calculations to the following maximally braided and maximally out-of-time-ordered $2k$-point function:
{\small
\begin{equation}
\begin{split}
&F_{2k}(t_1,\ldots,t_k;\sigma_1,\ldots,\sigma_k)  \\
&\equiv \frac{\big{\langle} X_1(t_1,\sigma_1) \rho^{-\frac{\varphi}{2\pi}} \, \left( \prod_{i=1}^{k-1} \big[X_{i+1}(t_{i+1},\sigma_{i+1})\rho^{-\frac{\varphi}{2\pi}},X_i(t_i,\sigma_i)\rho^{-\frac{\varphi}{2\pi}}\,\big] \right)\,X_k(t_k,\sigma_k) \rho^{-\frac{\varphi}{2\pi}} \, \rho^{\frac{2k\varphi}{2\pi}} \big{\rangle}}{\langle X_1X_1 \rangle \cdots \langle X_kX_k \rangle}
\end{split}
\label{eq:F2kdef}
\end{equation}
}%
which we studied in \cite{Haehl:2017pak} in the context of $AdS_2$ gravity. For simplicity, we will not analyze the full space of $(\tau_i,\sigma_i)$, but assume {\it a priori} that $t_1 > \ldots > t_k$ and $\sigma_1 < \ldots < \sigma_k$. One consequence of this assumption is that $\bar\epsilon$ never propagates and we can focus on the mode $\epsilon$ alone. We further assume that all operator dimensions are equal, and we set $\varphi = \delta$ and compute the leading term as $\delta \rightarrow 0$. 

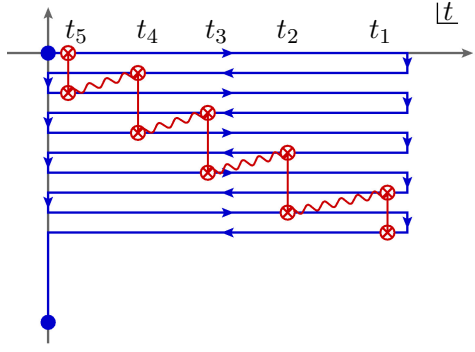
\begin{figure}
\begin{center}
\includegraphics[width=.4\textwidth]{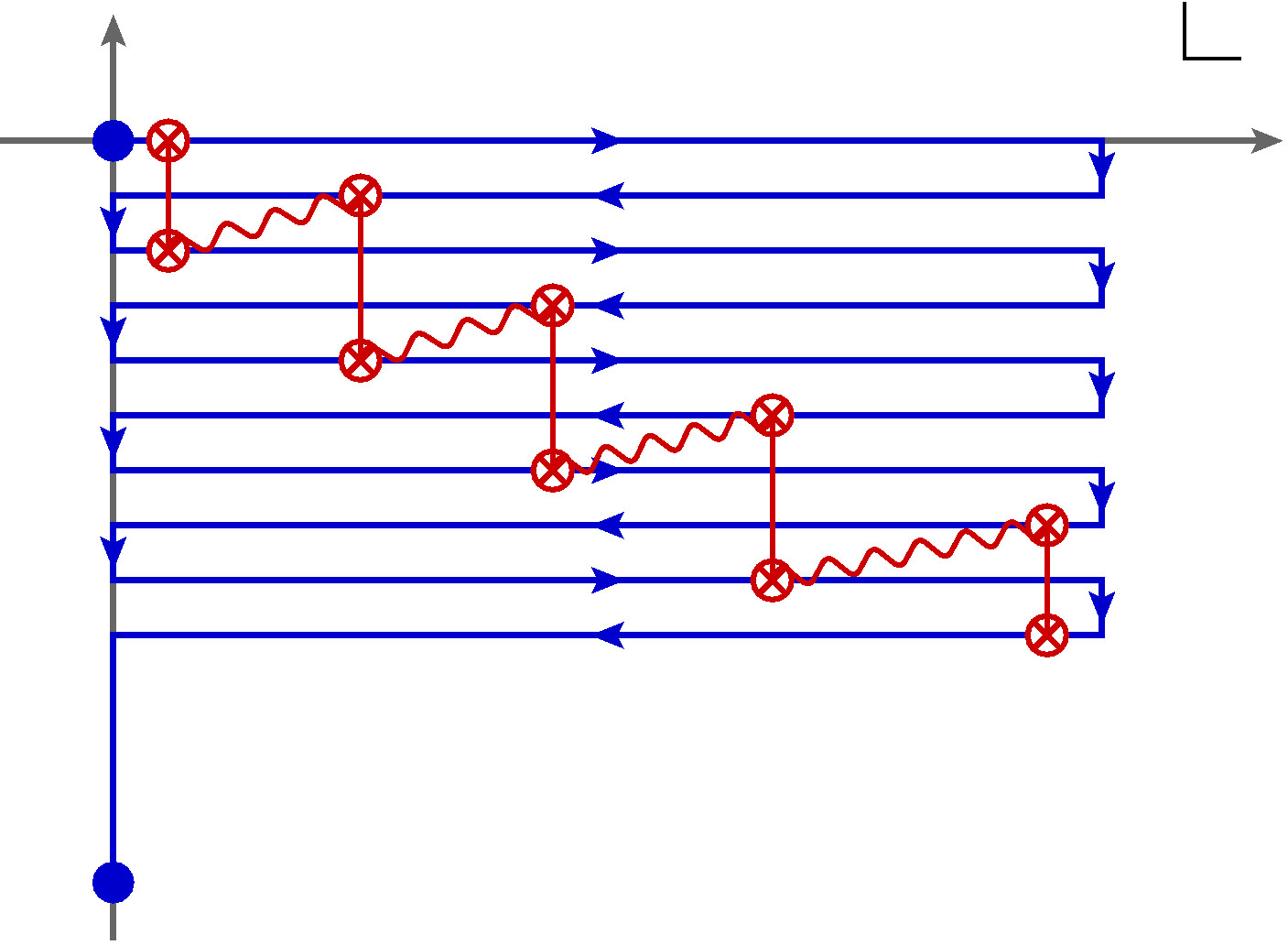}
\put(-12,123){$t$}
\put(-155,116){$t_5$}
\put(-128,116){$t_4$}
\put(-102,116){$t_3$}
\put(-75,116){$t_2$}
\put(-40,116){$t_1$}
\end{center}
\caption{Illustration of the arrangement of operators in our $2k$-point observable $F_{2k}$, for the case $k=5$. It provides a generalization of the familiar 4-point OTOC, which has the property that the associated time of exponential growth (which we call as $k$-scrambling time $t_*^{(k)}$) is maximal: $t_*^{(k)} = (k-1) \log c$. The red solid lines indicate bilocal operators ${\cal B}_{(h,\bar h)}$. The red wiggly lines schematically indicate the (maximally braided) propagation of the scrambling mode $\epsilon$.}
\label{fig:2kotoc}
\end{figure}

On the $k$-OTO contour, only the following terms contribute to the part of $F_{2k}$ that exhibits the longest exponential growth:
{\small
\begin{equation}
\begin{split}
& F_{2k} (t_1,\ldots,t_k;\sigma_1,\ldots,\sigma_k)\big{|}_{\text{exp.\ growing}}\\
& =  \big{\langle} {\cal T}_{{\cal C}_k} \; {\cal B}^{(1)2k,2k-2}_{(h,\bar h)}(t_1,\sigma_1,t_1,\sigma_1) \, \left( \prod_{i=1}^{k-2} {\cal B}^{(2)2k-2i+1,2k-2i-2}_{(h,\bar h)}(t_{i+1},\sigma_{i+1},t_{i+1},\sigma_{i+1}) \right)  \, {\cal B}^{(1)3,1}_{(h,\bar h)}(t_k,\sigma_k,t_k,\sigma_k) \big{\rangle}
\end{split}
\end{equation}
}\normalsize
This expression is illustrated in Fig.\ \ref{fig:2kotoc}.

In the small $\delta$ limit, we can make use of the following approximation:
{\small
\begin{equation}
\begin{split}
&{\cal B}^{(1)\alpha,\alpha-2}_{(h,\bar h)}(t_r,\sigma_r,t_r,\sigma_r) \rightarrow - \frac{h}{\delta} (\epsilon_r^\alpha - \epsilon_r^{\alpha-2})\,,
 \qquad\quad {\cal B}^{(2)\alpha,\alpha-3}_{(h,\bar h)}(t_{r},\sigma_{r},t_{r},\sigma_{r})\rightarrow \frac{(2h+1)h}{9\delta^2}\, (\epsilon_r^\alpha-\epsilon_r^{\alpha-3})^2
\end{split}
\end{equation}
}\normalsize
where we are also dropping all $\bar\epsilon$ dependence for the reason explained above. In this approximation, 
{\small
\begin{equation}
\label{eq:F2kBB}
 F_{2k}\big{|}_{\text{exp. gr.}} =\frac{(2h+1)^{k-2} h^k}{9^{k-2}\,\delta^{2(k-1)}}\,  \big{\langle} {\cal T}_{{\cal C}_k} \, (\epsilon_1^{2k} - \epsilon_1^{2k-2}) (\epsilon_2^{2k-1} - \epsilon_2^{2k-4})^2 \cdots (\epsilon_{k-1}^5 - \epsilon_{k-1}^2)^2 (\epsilon_k^3 - \epsilon_k^1) \big{\rangle} 
\end{equation}
}\normalsize
It is now a matter of simple combinatorics to evaluate this correlator to leading order in large central charge and small $\delta$. The $(2k-2)$-point function of $\epsilon$'s factorizes into two-point functions at leading order. These are of ${\cal O}(\delta^0)$ (as opposed to higher order in $\delta$) if and only if adjacent $\epsilon$'s in \eqref{eq:F2kBB} are ``braided'' along the contour. By this we just mean the following:
\begin{equation}
   \langle {\cal T}_{{\cal C}_k} \; \epsilon_r^{\alpha} (\epsilon_\ell^{\beta} - \epsilon_\ell^{\beta-\gamma}) \rangle = \left\{ 
   \begin{aligned} &\frac{12 \pi i}{c} \, e^{(t_r - t_\ell) - (\sigma_r - \sigma_\ell)} \qquad\quad (\alpha = \beta-1) \\
   & {\cal O}(\delta) \qquad\qquad\qquad\qquad\quad\;\;\; (\alpha > \beta) 
   \end{aligned} \right. 
\end{equation} 
for $\gamma=2,3$.\footnote{ This gives a completely Lorentzian meaning to the term ``maximally braided'', which originally referred to the arrangement of operators in Euclidean time. This property also shows that $2k$-point functions that are less braided (and less out-of-time-order) than $F_{2k}$ will generically have leading terms that are less divergent as $\delta \rightarrow 0$ and will have shorter associated timescales of exponential growth.}
By this reasoning, we can write 
{\small
\begin{equation}
\label{eq:F2kBB2}
\begin{split}
 F_{2k}\big{|}_{\text{exp. growing}} 
  &= \frac{2^{k-3} (-2\pi i)^{k-1}(2h+1)^{k-2} (2h)^k}{3^{k-3} \, \delta^{2k-2} }\; \times \, e^{t_\text{tot}- t_*^{(k)} - \sigma_\text{tot} } 
\end{split}
\end{equation}
}\normalsize
up to terms of order ${\cal O}(\delta^{-(2k-1)})$.
In the above, the total duration of the ``experiment'' is $t_{\text{tot}} \equiv t_1 - t_k$, the total ``extension'' is $\sigma_\text{tot} \equiv \sigma_k - \sigma_1$, and the {\it $k$-scrambling time} is defined as 
\begin{equation} 
\boxed{
t_*^{(k)} \equiv\frac{\beta}{2\pi} \,  (k-1) \, \log c \,.
}
\end{equation} 
For $k=3$, the above reproduces the very first term in \eqref{eq:6ptres}.

This result informs us that the correlator $F_{2k}$ keeps growing exponentially at least until a time $t_*^{(k)}$ whence it reaches an ${\cal O}(c^0)$ value. This $k$-scrambling time depends on $k$ because the Lyapunov exponent (i.e., the rate of the growth) is the same for all $k$. This interplay allows for the correlator to have a characteristic timescale of exponential growth which increases linearly with the number of operator insertions. This led us to the conclusion that the observable $F_{2k}$ measures genuinely new features of the scrambling process, which are more fine-grained in the sense that they take increasingly longer to saturate \cite{Haehl:2017pak}. From \eqref{eq:F2kBB2} we also see that additional spatial separation of the operators, $\sigma_\text{tot}$, tends to further increase this time scale. Note also that the butterfly velocity, just like the Lyapunov exponent, does not grow with $k$.

\vspace{10pt}
\section{Conclusions and Outlook}
\label{conclusions}

In this paper we have derived the effective theory of stress tensor soft modes in two-dimensional CFTs at large central charge. We have seen that the effective field theory bears strong similarity to the Schwarzian action which describes the low energy physics of the SYK model and $AdS_2$ gravity. Nevertheless, it differs from the Schwarzian theory in an interesting way and provides a higher dimensional framework. After developing real time methods in this context, we have demonstrated that the theory reproduces all early time signatures of chaos, including the out-of-time-ordered four-point function which grows exponentially over a characteristic scrambling time $t_* = \frac{\beta}{2\pi} \log c$. We also computed the more fine-grained $2k$-point functions discussed in \cite{Haehl:2017pak} and verified that they have longer characteristic scrambling times $t_*^{(k)} \sim (k-1) \, t_*$.

We now discuss various possible extensions of our work. The context of two-dimensional conformal field theory is special in that it is non-dissipative, thus the interesting potential connection between early-time chaos and late-time diffusion and transport cannot be probed in this context. One way to get to a more generic situation is by deforming the theory, either by a relevant operator (resulting in massive theory) or by the $T \bar{T}$ deformation and its variants \cite{Smirnov:2016lqw,Cavaglia:2016oda}. While the former set of deformations is more physical, the latter is solvable, at least for some quantities. Such deformations, perhaps studied in conformal perturbation theory, turn on genuine dissipative effects and simultaneously are expected to move us away from ``near-coherent'' maximal chaos. Hence these are interesting contexts to further study effective field theories of chaos. Generally, it will be important to study theories with non-maximal Lyapunov exponent in our framework and understand how the effective field theory description needs to be modified in that case.

Another issue that can be probed using such deformations is the bound on chaos \cite{Maldacena:2015waa}. Since the chaos exponent is maximal in our setup, any deformations of the theory impose a sign constraint on the shift of the Lyapunov exponent. It would be interesting to see what are those restrictions, which would be -- perhaps new -- manifestations of unitarity in quantum field theory. 

Most ambitiously, it is interesting to study higher-dimensional theories. If there is a shift symmetry responsible for chaos, it would suggest a new structure in hydrodynamical effective field theory, which would manifest such shift symmetry as a spontaneously and explicitly broken gauge symmetry. In our context, these features originated from conformal symmetry, where their microscopic origin is transparent. While one could first try to generalize our CFT arguments to higher dimensions, it remains to be understood what would be the origin of the structures postulated in \cite{Blake:2017ris} for non-conformal or not maximally chaotic theories.

We wish to draw the reader's attention to a feature of the Noether currents in section \ref{sec:noether}. As shown there, the Noether current for constant shifts of the soft modes can be interpreted as an entropy current. This is analogous to recent discussions of a $U(1)_T$ symmetry in hydrodynamics, whose associated current is the entropy current \cite{Haehl:2015pja,Haehl:2014zda,Haehl:2015foa}. The $U(1)_T$ symmetry emerges in coarse grained hydrodynamic states as a universal gauge redundancy. At least in the simple (non-dissipative) context of the present discussion, it is tempting to identify the corresponding part of $SL(2,\mathbb{R})$ with a version of $U(1)_T$. A rewriting of our effective action as a sigma model resembling the formal structure of hydrodynamic effective actions in a more obvious way (similar to \cite{Jensen:2016pah}) would presumably make this analogy more precise.

Moving away from our interest in early time chaos, we have written a perturbative effective field theory for the soft modes, which are the reparametrization modes of the 2-dimensional conformal field theory. 
Extending our effective field theory to higher orders should summarize the contribution of the identity block to correlation functions, at large central charge (see Appendix \ref{app:eucl3} for an example calculation of the third order $\epsilon\epsilon\epsilon$-vertex relevant for computing loop corrections). It would be interesting to phrase the simplifying limits of the identity conformal blocks in the present language -- those should correspond to re-summation of perturbation theory. Perhaps such a perspective can be used to identify other simplifying limits, using tools of effective field theory. We note that our effective field theory is inherently Lorentzian, thus such methods will not rely on analytic continuation from Euclidean space.


\vspace{20pt}
\section*{Acknowledgements}

We thanks Tarek Anous, Jan de Boer, Eliot Hijano, Matt Hodel, Kristan Jensen, Hong Liu, Henry Maxfield, Mukund Rangamani and Gabor Sarosi for useful conversations and correspondence. FH is grateful for support by the Simons grant ``It from Qubit'' and hospitality by the Mainz Institute for Theoretical Physics and GGI Florence where part of this work was done. The work of MR is supported by a Discovery grant from NSERC.

\appendix

\vspace{20pt}

\section{Euclidean Computations} 
\label{app:euclidean}

In this appendix we collect some details on the Euclidean calculations.

\subsection{Stress Tensor Two-Point Function}
\label{app:fourier}

We wish to do the Fourier transform of the energy-momentum two-point function more carefully. Consider
\begin{equation}
G^E_{(T)}(\omega_E,k) = \frac{c}{32}\int_{-\infty}^\infty d\sigma\int_0^{2\pi} d\tau \, \frac{e^{-i \omega_E \tau-i k \sigma}}{\sin (\frac{\tau+i\sigma}{2})^4}
\end{equation}
where we keep for the purpose of integration $\tau, \sigma$ to be independent complex variables. We change variables to $y=e^{i\tau}$, $r=|\sigma|$. This gives
\begin{equation}
G^E_{(T)}(\omega_E,k) = -\frac{c\,i}{2}\int_{0}^\infty dr \oint_{|y|=1} dy \;y^{1-\omega} \left( \frac{e^{(2-ik)r}}{(y-e^r)^4} + \frac{e^{-(2-ik)r}}{(y-e^{-r})^4} \right)
\end{equation}
We now perform the $y$-integral. This picks up residues from the second term in the bracket, but not from the first one. In addition, depending on the sign of $\omega$, we do or do not pick up poles from the factor $y^{1-\omega}$. Treating these two cases separately, we find that the answer can be summarized as follows:
%
\begin{equation}
\begin{split}
  G^E_{(T)}(\omega_E,k) &=  \frac{c\,\pi}{6}\, |\omega_E|(\omega_E^2-1)\, \int_{0}^\infty dr \, e^{-\text{sgn}(\omega_E)\,(\omega_E+ik)r}= \frac{c\,\pi}{6}\, \frac{\omega_E(\omega_E^2-1)}{\omega_E+ik} \,.
 \end{split}
\end{equation}

\vspace{10pt}
\subsection{Quadratic Action $W_2$}
\label{app:eucl}

Consider the quadratic action for the soft mode, eq.\ \eqref{quad} with $\langle T(z_1) T(z_2) \rangle = \frac{c/2}{(z_1-z_2)^4}$. We change coordinates according to 
\begin{equation}
 z_k= e^{-iu_k} \,, \qquad \bar{z}_k = e^{i\bar{u}_k} \qquad (k=1,2) \,.
\end{equation}
The sources transform as
\begin{equation}
  \bar{\partial}\epsilon(z_k,\bar{z}_k) \; \longrightarrow \; \frac{\partial z_k(u_k)}{\partial u_k}   \, \bar{\partial} \epsilon(u_k,\bar{u}_k)\,.
\end{equation}
We then obtain
\begin{equation}
\begin{split}
W_2 &= - \frac{c}{64}\int d^2u_1 d^2u_2 \, \frac{\bar{\partial} \epsilon(u_1,\bar{u}_1) \, \bar{\partial}\epsilon(u_2 , \bar{u}_2) }{\sin^4 \left( \frac{u_1-u_2}{2} \right) }  \\
&= - \frac{c}{64} \int d\tau_1 d\tau_2 d\sigma_1 d\sigma_2 \, \frac{(\partial_{\tau_1}+i \partial_{\sigma_1}) \epsilon(\tau_1,\sigma_1) \; (\partial_{\tau_2}+i \partial_{\sigma_2}) \epsilon(\tau_2,\sigma_2) }{\sin^4 \left( \frac{\tau_1-\tau_2 + i (\sigma_1 - \sigma_2)}{2} \right) } 
\end{split}
\end{equation}
where we set $u_k = \tau_k + i \sigma_k$ and $\bar{u}_k = \tau_k - i \sigma_k$ for $k=1,2$.

In order to perform perturbative calculations in the effective theory, we need the propagator of the soft mode. Using the Fourier transform of the energy-momentum 2-point function (i.e., the inverse of the calculation done in section \ref{app:fourier}), and integrating by parts, one has
\begin{equation}
\begin{split}
W_2
&=- \frac{c\,\pi}{12} \frac{1}{(2\pi)^2} \sum_{\omega_E} \int dk \int d\tau_1d\tau_2d\sigma_1d\sigma_2 \, (\partial_{\tau_1}+i \partial_{\sigma_1}) \epsilon(\tau_1,\sigma_1) \; (\partial_{\tau_2}+i \partial_{\sigma_2}) \epsilon(\tau_2,\sigma_2) \\
&\qquad\qquad\qquad\qquad\qquad\qquad\qquad\qquad\quad \times \frac{\omega_E(\omega_E^2-1)}{\omega_E+ik} \; e^{i\omega_E (\tau_1-\tau_2) + i k (\sigma_1-\sigma_2)} \\
&= -\frac{c\,\pi}{12} \frac{1}{(2\pi)^2}  \sum_{\omega_E} \int dk \int d\tau_1d\tau_2d\sigma_1d\sigma_2 \,  \epsilon(\tau_1,\sigma_1) \, \epsilon(\tau_2,\sigma_2) \\
&\qquad\qquad\qquad\qquad\qquad\qquad\qquad\qquad\quad \times \frac{\omega_E(\omega_E^2-1)}{\omega_E+ik} \; (\omega_E+ik)^2 \, e^{i\omega_E (\tau_1-\tau_2) + i k (\sigma_1-\sigma_2)} \\
&= -\frac{c\,\pi}{12} \frac{1}{(2\pi)^2}  \sum_{\omega_E} \int dk \,  \epsilon(\omega_E,k) \, \epsilon(-\omega_E,-k)   \,\omega_E\,(\omega_E^2-1) \, (\omega_E+ik) \,. \\
\end{split}
\label{eq:WquadCalc}
\end{equation}
When $k=0$ this is the linearized soft mode action used in \cite{Maldacena:2016upp}.
By inverting the integrand we can read off the propagator:
\begin{equation}
\langle \epsilon(\omega_E,k) \, \epsilon (-\omega_E, -k) \rangle=  -\frac{24\pi}{c} \frac{1}{\omega_E\,(\omega_E^2-1) \, (\omega_E+ik) } \,.
\end{equation}
The Fourier transform of this Euclidean propagator is
\begin{equation}
\begin{split}
 \langle \epsilon(\tau,\sigma)\epsilon(0,0) \rangle &= \frac{1}{(2\pi)^2} \, \sum_{\omega_E} \int dk \; \langle \epsilon(\omega_E,k) \, \epsilon (-\omega_E, -k) \rangle \; e^{i(\tau\omega + \sigma k)} \\
 &= \frac{12}{c} \bigg[ \Theta(\sigma) \left( \frac{1}{2} - 2 \, \sin^2\left( \frac{\tau+i\sigma}{2} \right) \, \log\big( 1 - e^{i(\tau+i\sigma)} \big) - \frac{3}{4} e^{i(\tau+i\sigma)} \right) \\
 &\quad\, - \Theta(-\sigma) \left( \frac{1}{2} - 2 \, \sin^2\left( \frac{\tau+i\sigma}{2} \right) \, \log \big( 1 - e^{-i(\tau+i\sigma)} \big) - \frac{3}{4} e^{-i(\tau+i\sigma)} \right)  \bigg]
\end{split}
\end{equation}

There is also an action for the mode $\bar\epsilon$ which differs from \eqref{eq:WquadCalc} only by the sign of $k$ in the last factor, leading to the propagator \eqref{eq:epsPropBar} for the mode $\bar\epsilon$. Finally, by Fourier transforming \eqref{eq:WquadCalc}  back to position space, we obtain the quadratic action in the form \eqref{eq:WquadCalc2}.

\vspace{10pt}
\subsection{Cubic action $W_3$}
\label{app:eucl3}

It is easy to generalize our analysis of the soft mode propagator (encoded in the quadratic action) to third order (and higher orders). Indeed, the cubic vertex for the soft mode is generated by the stress tensor three-point function \cite{Osborn:1993cr}:
\begin{equation}
\label{eq:Wcubic}
\begin{split}
W_{3}&= \frac{1}{3!}\int d^2 z_1 \, d^2 z_2 \, d^2 z_3 \, \bar{\partial} \epsilon(z_1,\bar{z}_1) \,  \bar{\partial} \epsilon(z_2,\bar{z}_2) \,  \bar{\partial} \epsilon(z_3,\bar{z}_3) \, \langle T(z_1) T(z_2) T(z_3) \rangle_c  \\
&= 
 \frac{1}{3!}\int d^2 z_1 \, d^2 z_2 \, d^2 z_3 \, \bar{\partial} \epsilon(z_1,\bar{z}_1) \,  \bar{\partial} \epsilon(z_2,\bar{z}_2) \,  \bar{\partial} \epsilon(z_3,\bar{z}_3) \,\frac{c}{(z_1-z_2)^2(z_2-z_3)^2(z_3-z_1)^2} \\
&=   \frac{c}{384} \int d\tau_1d\sigma_1d\tau_2d\sigma_2d\tau_3d\sigma_3 \, \frac{(\partial_{\tau_1}+i\partial_{\sigma_1}) \epsilon(\tau_1,\sigma_1) \;(\partial_{\tau_2}+i\partial_{\sigma_2}) \epsilon(\tau_2,\sigma_2)\;(\partial_{\tau_3}+i\partial_{\sigma_3}) \epsilon(\tau_3,\sigma_3)}{\sin\left(\frac{\tau_{12}+i\sigma_{12}}{2}\right)^2 \, \sin\left(\frac{\tau_{23}+i\sigma_{23}}{2}\right)^2 \, \sin\left(\frac{\tau_{31}+i\sigma_{31}}{2}\right)^2 }
\end{split}
\end{equation}
We use the following Fourier identity: 
\begin{equation}
  \frac{1}{\sin\left( \frac{\tau+ i \sigma}{2} \right)^2} = -8\pi\, \frac{1}{(2\pi)^2} \sum_{\omega_E} \int dk \, \frac{\omega_E}{\omega_E+ik} \, e^{-i(\omega_E \tau  + k \sigma)} \,.
\end{equation}
We can then write $W_3$ as 
{\small
\begin{equation}
\label{eq:Wcubic}
\begin{split}
W_{3}&=
\frac{c\,(8\pi i)^3}{384} \int d\tau_1d\sigma_1d\tau_2d\sigma_2d\tau_3d\sigma_3 \, \frac{1}{(2\pi)^6} \sum_{\omega_{E}^{12},\,\omega_E^{23},\,\omega_E^{31}} \int dk^{12} dk^{23}dk^{31} \,\epsilon(\tau_1,\sigma_1) \, \epsilon(\tau_2,\sigma_2) \, \epsilon(\tau_3,\sigma_3) \\
&\quad \times \frac{\omega_E^{12}\, \omega_E^{23}\, \omega_E^{31} \, [(\omega_E^{31}-\omega_E^{12}) + i (k^{31}-k^{12})] \, [(\omega_E^{12}-\omega_E^{23}) + i (k^{12}-k^{23})] \, [(\omega_E^{23}-\omega_E^{31}) + i (k^{23}-k^{31})]}{(\omega_E^{12}+i k^{12})(\omega_E^{23}+i k^{23})(\omega_E^{31}+i k^{31})} \\
&\quad \times e^{-i \left( \omega_E^{12}\tau_{12} + k^{12} \sigma_{12} +  \omega_E^{23}\tau_{23} + k^{23} \sigma_{23}  +  \omega_E^{31}\tau_{31} + k^{31} \sigma_{31}  \right)} 
\\
& = 
\frac{c\,(8\pi i)^3}{384}\frac{1}{(2\pi)^6} \sum_{\omega_{E}^{12},\,\omega_E^{23},\,\omega_E^{31}} \int dk^{12} dk^{23}dk^{31} \\
&\quad \times \frac{\omega_E^{12}\, \omega_E^{23}\, \omega_E^{31} \, [(\omega_E^{31}-\omega_E^{12}) + i (k^{31}-k^{12})] \, [(\omega_E^{12}-\omega_E^{23}) + i (k^{12}-k^{23})] \, [(\omega_E^{23}-\omega_E^{31}) + i (k^{23}-k^{31})]}{(\omega_E^{12}+i k^{12})(\omega_E^{23}+i k^{23})(\omega_E^{31}+i k^{31})} \\
&\quad \times \epsilon(\omega_E^{31}-\omega_E^{12},k^{31}-k^{12}) \, \epsilon(\omega_E^{12}-\omega_E^{23},k^{12}-k^{23}) \, \epsilon(\omega_E^{23}-\omega_E^{31},k^{23}-k^{31}) \\
& = 
\frac{c\,(8\pi i)^3}{384}\frac{1}{(2\pi)^6} \sum_{\omega_{E}^{12},\,\omega_E^{23},\,\omega_E^{31}} \int dk^{12} dk^{23}dk^{31} \;\epsilon(\omega_E^{31},k^{31}) \, \epsilon(-\omega_E^{23},-k^{23}) \, \epsilon(\omega_E^{23}-\omega_E^{31},k^{23}-k^{31}) \\
&\quad \times \frac{\omega_E^{12}\, (\omega_E^{23}+\omega_E^{12})\, (\omega_E^{31}+\omega_E^{12}) \, (\omega_E^{31} + i k^{31}) \, (-\omega_E^{23} - i k^{23}) \, [(\omega_E^{23}-\omega_E^{31}) + i (k^{23}-k^{31})]}{(\omega_E^{12}+i k^{12})[\omega_E^{23}+\omega_E^{12}+i (k^{23}+k^{12})][\omega_E^{31}+\omega_E^{12}+i (k^{31}+k^{12})]} 
\end{split}
\end{equation}
}\normalsize
where in the last step we shifted variables as $\{\omega_E^{31},\, \omega_E^{23}\} \mapsto \{\omega_E^{31} + \omega_E^{12},\,\omega_E^{23}+\omega_E^{12}\}$ and similarly for $k$-variables.
We can now perform the sum over $\omega_E^{12}$ and the integral over $k^{12}$ explicitly (this is a consequence of energy conservation). This gives:
{\small
\begin{equation}
\label{eq:Wcubic}
\begin{split}
W_{3}
& = 
\frac{ic}{144\pi^2} \sum_{\omega_E^1,\,\omega_E^b,\,\omega_E^3} \int dk^1dk^2dk^3 \, \epsilon(\omega_E^1,k^1) \, \epsilon(\omega_E^2,k^2) \, \epsilon(\omega_E^3,k^3)  \\
&\quad \times\delta(\omega_E^1+\omega_E^2+\omega_E^3)\,\delta(k^1+k^2+k^3)\, \frac{1}{3!}\sum_{\substack{a,b=1,2,3\\ a\neq b}}(\omega_E^a+i k^a) (2\omega_E^a+\omega_E^b)\omega_E^b\left((\omega_E^b)^2-1\right)
\end{split}
\end{equation}
}\normalsize
From this we can read off the $\langle \epsilon(\omega_E^1,k^1)\,\epsilon(\omega_E^2,k^2)\, \epsilon(\omega_E^3,k^3) \rangle$-vertex, which is just the integration kernel of \eqref{eq:Wcubic}.
This vertex will be useful for computing quantum (loop) corrections in the soft mode exchange of the out-of-time-ordered correlators. We leave this as an interesting future task.

In position space, the Euclidean cubic action \eqref{eq:Wcubic} can be written as follows (with $\bar\partial = \frac{1}{2} (\partial_\tau + i \partial_\sigma)$):
\begin{equation}
\label{eq:WcubicPos}
\begin{split}
W_{3}
& = 
\frac{c}{18}\,(2\pi)^2 \int d\tau d\sigma  \; \bar{\partial} \epsilon(\tau,\sigma)\; \left[ (\partial_\tau^2 +1) \partial_\tau^2 \epsilon(\tau,\sigma)\; \epsilon(\tau,\sigma) + 2  (\partial_\tau^2 +1) \partial_\tau \epsilon(\tau,\sigma)\; \partial_\tau\epsilon(\tau,\sigma) \right] \,.
\end{split}
\end{equation}

\vspace{10pt}
\section{$SL(2,\mathbb{R})$ Generators}
\label{sl2}


We saw that each chiral part of the Euclidean quadratic action \eqref{eq:WquadCalc2} has three $SL(2,\mathbb{R})$ zero modes, and also the coupling to matter respects these symmetries. Those transformations correspond to holomorphic reparametrizations which preserve the saddle point solution, even when their parameter is taken to be a general function $\epsilon(z, \bar{z})$. Those modes have to be treated as gauge redundancies, as in \cite{Maldacena:2016upp}. 

We start by working on the Euclidean plane. An infinitesimal vector field of the form $v \equiv v^z(z) \, \partial_z + v^{\bar z}(\bar z) \, \partial_{\bar z} \equiv v \, \partial + \bar v \, \bar\partial$ induces an infinitesimal conformal transformation $(z,\bar z) \mapsto  (z -  v , \, \bar z -  \bar{v})$, which we represent on a primary $\phi(z,\bar z)$ of dimension $(h,\bar h)$ as 
\begin{equation}
\begin{split}
  \phi(z,\bar z) \mapsto \phi'(z,\bar z) 
   &= -  \big( h\,\partial v + \bar h \,\bar\partial \bar v + v\partial + \bar v \bar \partial \big) \phi(z,\bar z)  \,.
 \end{split}
\end{equation}

Acting on the plane parametrized by $z$, the $SL(2,\mathbb{R})$ generators $L_n$ can be represented as vector fields ${\cal L}_n = z^{n+1} \partial_z$, such that they generate the conformal transformations
\begin{equation}
  L_n \phi = -  z^{n+1} \partial \phi -h(n+1) z^n \, \phi \,,\qquad
  {\bar L}_n \phi = -  \bar{z}^{n+1} \bar\partial \phi - \bar{h}(n+1) {\bar z}^n \, \phi \,.
\end{equation}

If we now go to (Euclidean) thermal coordinates via $(z,\bar z) = (e^{-iu},e^{i\bar{u}}) = (e^{-i(\tau+i\sigma)},e^{i(\tau-i\sigma)})$, the generators understood as vector fields take the form
\begin{equation}
\label{eq:Lnvec}
   {\cal L}_n  = ie^{-in u} \partial_u = \frac{i}{2} \, e^{-in (\tau+i\sigma)} \, (\partial_\tau - i \partial_\sigma) \,,\qquad
   \bar{\cal L}_n = -ie^{in\bar u} \partial_{\bar u}  =- \frac{i}{2} \, e^{in (\tau-i\sigma)} \, (\partial_\tau + i \partial_\sigma) \,.
\end{equation}
They induce a transformation on dimension $(h,\bar h)$ primaries as 
\begin{equation}
  L_n \phi = - \frac{i}{2} \, e^{-in(\tau+i\sigma)} \, \big(\partial_\tau - i \partial_\sigma - 2 i n h\big)\, \phi \,,\qquad
  {\bar L}_n \phi = \frac{i}{2} \, e^{in(\tau-i\sigma)} \, \big(\partial_\tau + i \partial_\sigma + 2in\bar{h}\big)\, \phi \,.
\end{equation}

We now analytically continue $\tau \rightarrow t_E + i t$, where $t$ is the Lorentzian time. Then each segment of a multi-segment contour, for example the Schwinger-Keldysh contour, can be placed in constant Euclidean time $t_E=\varphi$. We can choose the convention that all segments of a contour are equally separated in Euclidean time, i.e. for a $k$-OTO contour which contains $2 k$ segments, the $\ell$'th segment has time variables $t -i\,\frac{2 \pi (\ell-1)}{2k}$.\footnote{  Note that these real times correspond to the usual picture of the Schwinger-Keldysh (or $k$-OTO) contour in the complex time plane, where real time evolution proceeds forwards and backwards parallel to the real line, and Euclidean evolution runs in the negative imaginary direction. Forward and backward segments of the contour are separated by $-\frac{\pi i}{k}$.} The fields on each such segment are then functions of a single Lorentzian time, which is obtained by the above Wick rotation. The generators with respect to Lorentzian time are then
\begin{equation}
  L_n \phi = -\frac{1}{2} \, e^{-in\varphi} \, e^{n(t+\sigma)} \, \big(\partial_t + \partial_\sigma + 2n h\big)\, \phi \,,\qquad
  {\bar L}_n \phi = \frac{1}{2} \, e^{in\varphi} \, e^{-n(t-\sigma)}  \, \big(\partial_t - \partial_\sigma - 2n\bar{h}\big)\, \phi \,.
\end{equation}

If we think of our soft mode $\epsilon$ as a field that transforms like a primary of dimension $(-1,0)$ under $SL(2,\mathbb{R})$, we have in particular 
\begin{equation}
\begin{split}
  L_1\, \epsilon &= -\frac{1}{2} \, e^{-i\varphi} \, e^{t+\sigma} \, \big(\partial_t + \partial_\sigma -2\big)\,\epsilon\,,\\
  L_0\, \epsilon &= -\frac{1}{2} \,  \big(\partial_t + \partial_\sigma \big)\, \epsilon\,,\\
  L_{-1} \,\epsilon &= -\frac{1}{2} \, e^{i\varphi} \, e^{-(t+\sigma)} \, \big(\partial_t + \partial_\sigma+2 \big)\epsilon\,.
  \end{split}
\end{equation}
Thus, the holomorphic exponentially growing mode $\epsilon \sim e^{t+\sigma}$ is annihilated by $L_1$. Similarly, the anti-holomorphic exponentially growing mode $\bar \epsilon \sim e^{t - \sigma}$ is annihilated by $\bar{L}_1$. 
The zero modes which are not exponentially growing ($\epsilon \sim e^{-(t+\sigma)}$ and $\epsilon \sim \text{const.}$) are likewise annihilated by $L_{-1}$ and $L_0$ (and similarly for $\bar\epsilon$).
We see that the shift symmetry, annihilating the exponentially growing mode, comes about naturally from the microscopic conformal field theory: it is associated with gauge redundancies due to $SL(2,\mathbb{R})$ transformations which are a relict of the conformal symmetry which is explicitly and spontaneously broken.

\vspace{10pt}
\bibliographystyle{JHEP}

\providecommand{\href}[2]{#2}\begingroup\raggedright\endgroup


\end{document}